%% file: main.tex
\documentclass[conference]{IEEEtran}
\IEEEoverridecommandlockouts

\usepackage{cite}
\usepackage{amsmath,amssymb}
\usepackage{graphicx}
\usepackage{booktabs,multirow,makecell,array,tabularx}
\usepackage{enumitem}
\usepackage{pifont}
\usepackage{xspace}
\usepackage{listings}
\usepackage[hidelinks]{hyperref}

\newcommand{\Code}[1]{\lstinline{#1}}
\newcommand{\qubic}{\textsl{QubiC}\xspace~2.0\xspace}
\newcommand{\sys}{\mbox{\textsl{Ant-Q}}\xspace}
\newcommand{\eg}{\textit{e.g.}}
\newcommand{\ie}{\textit{i.e.}}
\newcommand{\ack}[1]{\section*{Acknowledgment}#1}

\begin{document}

\title{Breaking Memory Bottlenecks in Quantum Control Systems for Deeper, Longer Experiments and Higher-Throughput Computing}

\author{\IEEEauthorblockN{Yicheng Guang\textsuperscript{1,2}, Neel Vora\textsuperscript{2}, Yilun Xu\textsuperscript{2}, Yueqi Chen\textsuperscript{1} and Gang Huang\textsuperscript{2}}
\IEEEauthorblockA{\textsuperscript{1}\textit{University of Colorado Boulder}, Boulder, CO, USA \\
\textsuperscript{2}\textit{Lawrence Berkeley National Laboratory}, Berkeley, CA, USA \\
\{yicheng.guang, yueqi.chen\}@colorado.edu, \{yguang, nrvora, yilunxu, ghuang\}@lbl.gov}}

\maketitle

\input{sections/0Abstract}


\input{sections/intro}
\input{sections/background}
\input{sections/design}
\input{sections/implementation}
\input{sections/evaluation}
\input{sections/discuss}
\input{sections/related}
\input{sections/conclude}

\appendices
\input{sections/appendix}

\input{sections/Acknowledgement}

\section*{Data Availability}
The raw data, measurement and analysis scripts, benchmark circuits and bitstreams that support the findings of this study are openly available in the Ant-Q repository~\cite{antq} (release v1.11, DOI \href{https://doi.org/10.5281/zenodo.23250200}{10.5281/zenodo.23250200}). The archive pins the commit of every repository and lists the SHA-256 checksums of the bitstreams, and its README gives the command that rebuilds each table and figure. The Ant-Q gateware and software are part of the \textsl{QubiC} repositories~\cite{qubic_gateware, qubic_software, qubic_distproc}.

\bibliographystyle{IEEEtran}
\bibliography{ref}

\end{document}

%% file: sections/0Abstract.tex
\begin{abstract}

As quantum computing matures, there is an increasing need for deeper and longer experiments to characterize quantum devices precisely and higher circuit throughput to validate more domain applications.
However, this need is hindered by a memory bottleneck at the quantum control system layer, arising from the limited capacity of on-chip block random-access memory (BRAM) and the non-deterministic latency of dynamic random-access memory (DRAM).
To break this bottleneck, we present \sys, a memory hierarchy design for the \qubic control system that integrates DRAM with BRAM to support pipelined quantum circuit execution while keeping inter-circuit timing deterministic whenever each circuit outlasts the refill of the next.
A switch between command buffers then takes a constant $134~\text{ns}$ and, inside a circuit, acts on the qubit as a $144~\text{ns}$ idle.
On the bench, \sys runs deep one- and two-qubit randomized benchmarking beyond the command capacity of the baseline.
For circuits with at least $1000$ shots, it reduces the time the board spends beyond the circuit execution from $18.91\%$–$104.54\%$ to at most $0.53\%$ of the execution time, and it improves the board-level throughput of $30$-circuit batches by $51.06\%$ against a baseline with the same shared tables ($59.90\%$ against the stock baseline).
On a transmon qubit, \sys executes a $5101$-Clifford sequence and acquires uninterrupted $100~\text{s}$ Ramsey records with $0.01~\text{Hz}$ frequency spacing.
In simulations, these records detect $1/f$ frequency noise of $1.37$ and $1.62$ times lower spectral density than the same records split into the $2.048~\text{s}$ runs of the baseline.
The measured spectra resolve no $1/f$ component and match the measurement floor from $0.1$ to $100~\text{Hz}$.
\sys has been merged into the open-source \textsl{QubiC} gateware and software.

\end{abstract}

%% file: sections/intro.tex
\section{Introduction}

As quantum computing develops toward fault tolerance, the need for more precise characterization of qubit noise and gate fidelity continues to grow~\cite{berritta2026real,yan2012noise,riste2013charge,tornow2022restless,hughes2025trappedion,vion2002quantronium,magesan2012interleaved}.
Meanwhile, users from more domains explore the potential of quantum advantage, which pushes quantum computing systems toward higher throughput.

\begin{figure}[t]
    \centering
    \includegraphics[width=\columnwidth]{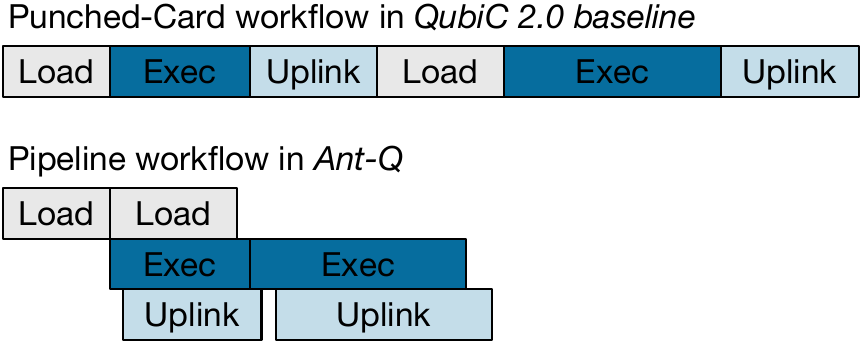}
    \caption{\sys transforms the workflow of the \qubic control system. In the pipeline workflow of \sys, circuit loading, execution and readout uplink are decoupled and overlap in time. As a result, readout data are uplinked shortly after execution begins.}
    \label{fig:workflow}
\end{figure}

Both needs fall on the quantum control system, the layer of the computing stack that bridges the classical and quantum worlds.
Figure~\ref{fig:workflow} shows the workflow of the \qubic baseline~\cite{xu2023qubic}, which resembles that of early punched-card classical computers.
The host first loads a circuit into the control board (downlink), and the board executes it.
Only after all shots complete does the board send the readout results back to the host (uplink), and only then is it ready for the next circuit.
This punched-card workflow assumes an upper bound on circuit size, so it cannot run the deep circuits and the long, high-sampling-rate experiments that precise characterization needs.
It also limits throughput, because the quantum processing unit (QPU) idles while circuits are loaded and readout data are transferred.

The punched-card workflow follows from a memory bottleneck of the control board.
Because qubits decohere quickly, the control system must operate in real time and meet strict deterministic timing constraints.
On the control board, only block random-access memory (BRAM) meets these constraints, with its single-cycle access latency.
However, BRAM capacity on control boards is highly limited, and adding BRAM increases routing complexity and fan-out, which reduces the achievable clock frequency~\cite{gungor2022optimizing}.
Under these constraints, the punched-card workflow becomes a natural choice.

In this work, we develop \sys, a memory-hierarchy system for the \qubic control board that breaks this bottleneck and turns the punched-card workflow into a pipeline (Fig.~\ref{fig:workflow}).
The key idea is to use the dynamic random-access memory (DRAM) on the same board as the main memory pool and to repurpose BRAM as an explicitly managed staging buffer.
On top of this hierarchy, \sys streams commands from DRAM, buffers and streams the readout, hands over the command buffers of all cores together and manages batches of circuits.
As a result, circuit loading, circuit execution and readout uplink run on separate data paths and overlap in time, like the coordinated work of an ant colony.

DRAM brings capacity but not determinism.
Its refresh cycles and row or bank conflicts can block an access for hundreds of nanoseconds~\cite{jedec_ddr4}.
To hide this latency, \sys places ping-pong buffers in BRAM between DRAM and the DSP cores.
When the next half of a buffer is ready, the switch between the two halves takes a constant $134~\text{ns}$.
Of this time, $120~\text{ns}$ is the restart of the DSP core, which the \qubic baseline also incurs between every two shots, and only $14~\text{ns}$ comes from \sys.
\sys also manages circuits at a fine grain and moves them in batches by Advanced eXtensible Interface (AXI) direct memory access (DMA)~\cite{amd_pg021}, which shortens circuit loading and readout uplink.
Because the readout now streams while a circuit runs, \sys can terminate a circuit once noise dominates its results and return the remaining time to the QPU.

\sys is implemented on the ZCU216 evaluation board~\cite{amd_ug1390} and has been merged into the \texttt{feat/ddr\_mem} branch of the \textsl{QubiC} gateware and software repositories~\cite{qubic_gateware, qubic_software}.
To evaluate \sys, we construct a benchmark suite of four physical experiments and $20$ computational tasks.
The four experiments are infeasible under the \qubic baseline, and \sys runs all of them, including Repeated Ramsey experiments with up to $5 \times 10^{4}$ shots~\cite{yan2012noise}.
For circuits with at least $1000$ shots, \sys reduces the time the control board spends beyond the circuit execution, which on the \qubic baseline goes mostly to circuit loading and readout uplink, from $18.91\%$–$104.54\%$ to at most $0.53\%$ of the execution time.
It improves the throughput of batches of $30$ heterogeneous circuits by $51.06\%$ on average against a baseline with the same shared tables and by $59.90\%$ against the stock baseline.
We further validate \sys on a superconducting transmon qubit~\cite{koch2007transmon} at the Advanced Quantum Testbed~\cite{aqt}.
There it executes a $5101$-Clifford randomized benchmarking (RB) sequence~\cite{knill2008randomized,magesan2011scalable} that the \qubic baseline cannot hold and gives Ramsey and RB results equivalent to the baseline within pre-registered margins~\cite{antq}.
Its uninterrupted $100~\text{s}$ Ramsey records have noise spectra consistent with the measurement floor from $0.1$ to $100~\text{Hz}$, with no resolved $1/f$ component.

DRAM already holds control sequences in other systems.
ARTIQ replays event sequences recorded in DRAM by DMA~\cite{artiq_dma}, and a control system for trapped ions on a radio-frequency system-on-chip (RFSoC) moves gate sequences from DRAM to the programmable logic by scatter-gather DMA~\cite{dudley2025sgdma}.
QICK streams accumulated readout buffers to the host while a program runs~\cite{qick_streamer}.
What \sys adds is the combination on \qubic, including commands streamed from DRAM to all DSP cores with a synchronized switch of constant cost between command buffers, the phase state carried across program segments, readout streamed in the same run, and batches of circuits with shared tables, together with their measured effect on a qubit and on throughput (Table~\ref{tab:compare}).
Two constraints shape this combination.
The DSP cores run in lock step, so all their command buffers must switch at one boundary at a known cost, and the oscillator phases must continue across program segments.
The approach carries over to other controllers that execute commands from on-chip memory, first stage the commands of each core in a ping-pong buffer filled from DRAM, then switch all cores together, and finally drain the readout into a ring buffer that the processor reads while the program runs.

In summary, this paper makes the following contributions.
\begin{itemize}
\item We trace the punched-card workflow of BRAM-based control boards to their memory bottleneck and identify three challenges of bringing DRAM into the real-time control path (Section~\ref{sec:background}).
\item We design \sys, a memory hierarchy that streams commands from DRAM through ping-pong buffers in BRAM, switches the buffers of all cores together in a constant $134~\text{ns}$ and streams the readout while the program runs (Section~\ref{sec:design}). \sys has been merged into the open-source \textsl{QubiC} gateware and software.
\item We evaluate \sys on the bench and on a transmon qubit (Section~\ref{sec:eval}). It runs a $5101$-Clifford RB sequence that the \qubic baseline cannot hold, reduces the time beyond the execution to at most $0.53\%$ for circuits with at least $1000$ shots, and raises the throughput of $30$-circuit batches by $51.06\%$ against a baseline with the same shared tables.
\end{itemize}

Section~\ref{sec:background} describes the demands on the control system and the memory bottleneck of the \qubic baseline, Section~\ref{sec:design} the design of \sys and Section~\ref{sec:implementation} its implementation.
Section~\ref{sec:eval} evaluates \sys on the bench and on a qubit, and Sections~\ref{sec:future_work} and~\ref{sec:related} discuss its limitations and related work.
The appendices give the details of the measurements and analyses.
The benchmark circuits, raw data and analysis scripts are released in~\cite{antq}.

%% file: sections/background.tex
\section{Problem Description}
\label{sec:background}
In this section, we first describe what deep, long-running experiments and high-throughput computing demand from the control system.
Then, we explain why the \qubic baseline cannot meet these demands.
Finally, we derive the challenges of bringing DRAM into the real-time control path, which Section~\ref{sec:design} takes up.

\subsection{Demands for Deep and Long-Running Experiments}
Precise characterization of a qubit relies on two kinds of experiments, and each kind strains the control system in its own way.

Long-duration, high-sampling-rate experiments probe environmental noise.
The single-qubit Ramsey sequences used for $1/f$ noise spectroscopy~\cite{yan2012noise} and charge-parity monitoring~\cite{riste2013charge} need both a large number of shots and strictly deterministic inter-shot timing.
For instance, capturing charge-parity fluctuations can require a measurement every $6~\mu\text{s}$ for thousands of samples.
The readout data of such experiments accumulate rapidly, so the control system must offload them to the host machine in time.
Otherwise, memory exhaustion splits the experiment into multiple runs, which breaks the deterministic inter-shot timing and can miss transient noise events.

Randomized benchmarking (RB)~\cite{knill2008randomized,magesan2011scalable}, which validates gate fidelity~\cite{hughes2025trappedion}, needs deep randomized circuits.
RB relies on error amplification.
The error of a single high-fidelity gate is too small to distinguish from measurement noise, so RB accumulates it over a deep sequence of gates until the survival probability decays measurably.
RB sequences reach a depth of $5101$ Clifford gates to achieve the necessary statistical resolution~\cite{tornow2022restless}.
These sequences consist of unique, randomly generated gates and cannot be rolled into shorter circuits by loop iteration, so the control system must store all their commands.

\subsection{Demands for Higher Throughput Computing}
Besides running experiments, the control system is a critical layer in the computing stack.
To raise system throughput and serve the jobs of more users, this layer must not stall the QPU.

In the current Noisy Intermediate-Scale Quantum (NISQ) era~\cite{preskill2018nisq}, most computational circuits are short, and their execution accounts for only a small fraction of the overall runtime.
A study of IBM's quantum systems~\cite{ravi2022quantum} defines the system overhead as the total time spent on queuing, circuit loading and readout transfer.
It reports that the median ratio of this overhead to the QPU execution time is about $10$, and $100$ or more for $25\%$ of jobs.
Queuing lies outside the control system.
During circuit loading and readout transfer, however, the QPU itself waits for the classical part to finish.

Noise wastes QPU time in another way.
On noisy hardware, gate errors and decoherence can turn the results of circuits validated against known classical results into distributions far from the ideal one~\cite{rethinking_noise}.
Executing the rest of the shot budget then wastes QPU time.
Recent works reduce the required number of shots at the software level~\cite{adaptive_fidelity_qufid}, but control systems such as the \qubic baseline provide no mechanism to terminate a running circuit once noise dominates its results.

\subsection{Quantum Control Systems}
To see why the \qubic baseline cannot meet these demands, we first describe how modern quantum control systems work.
Modern quantum control systems, such as \textsl{QubiC}~\cite{xu2023qubic} and the Quantum Instrumentation Control Kit (\textsl{QICK})~\cite{stefanazzi2022qick}, have largely moved from arbitrary waveform generators (AWGs) to field-programmable gate arrays (FPGAs).
FPGAs offer lower cost, real-time feedback loops, scalable qubit control channels and programmable control logic.
These systems partition the functionality between a Processing System (PS) and a Programmable Logic (PL).
The PS runs a general-purpose operating system such as Linux and handles the tasks without strict timing, for example receiving circuits from a remote host and returning readout results over TCP/IP.
The PL executes the circuit commands on a deterministic schedule and generates and captures pulses at precise time points with digital signal processing (DSP) cores.
The \qubic baseline cannot meet the demands above, for two reasons.

The first reason is its punched-card workflow.
In this workflow, the host first compiles a circuit and submits it to the PS.
The PS then downlinks the circuit to the PL and triggers its execution.
During execution, the PL stores the readout data and uplinks them to the PS in one batch after the circuit completes.
This workflow incurs a large classical overhead for short circuits.
On the $20$ benchmark circuits of Section~\ref{sec:eval-rq2}, the baseline runs $18.91\%$ to $313.04\%$ longer than the shots of each circuit alone.
The workflow also rules out early termination of noise-overwhelmed circuits, since the host sees the readout data only after all shots have finished.
Turning this workflow into a pipeline is difficult because the PS does not run deterministically.
If the PS does not downlink the circuit commands in time, the DSP stalls, the QPU stalls with it, and the results become incorrect.

The second reason is the memory bottleneck.
To meet the strict timing requirements, the PL stores circuits and readout results in BRAM, which provides single-cycle access latency and suffers no external bus contention.
BRAM is scarce, however.
The ZCU216, the deployment board of \textsl{QubiC 2.0}, offers only $4.75~\text{MB}$ of BRAM~\cite{amd_ug1390,amd_ds889}.
Long-duration, high-sampling-rate experiments exhaust this capacity with their readout data, and deep randomized circuits exceed it with their commands.
After envelope waveforms and auxiliary data, the default configuration of \qubic executes at most $2048$ commands, stores up to $1024$ readout records per qubit and supports no more than $14$ qubits.
For example, a single-qubit RB sequence with $1024$ Clifford gates already exceeds $2048$ commands, and \qubic rejects it (Section~\ref{sec:eval-device}).
Such experiments therefore need other instruments or customized FPGA firmware.
Adding BRAM does not remove the bottleneck, because the extra routing complexity and fan-out reduce the achievable clock frequency~\cite{gungor2022optimizing}.

\subsection{Challenges of Bringing DRAM into the Control Path}
\label{sec:challenges}
The control board already carries a much larger memory, its DRAM.
DRAM on the PL offers up to $4~\text{GB}$ on the ZCU216 and runs at a clock frequency close to that of the DSP and BRAM.
Its worst-case latency is also much lower than that of the PS, which can reach $1~\text{s}$~\cite{chen2025principled}.
DRAM is therefore the natural place for deep circuits and long readout records.
Bringing it into the real-time control path, however, raises three challenges.

\textbf{C1, non-deterministic latency.}
DRAM needs periodic refresh cycles to retain its data and can run into row or bank conflicts~\cite{jedec_ddr4}.
These operations can block an access for hundreds of nanoseconds, while a DSP core consumes its commands on a fixed schedule.

\textbf{C2, lock-step cores.}
The DSP cores run in lock step.
When the commands of a deep circuit arrive in segments, all command buffers must switch at one boundary at a known cost, and the oscillator phases must continue across the segments.

\textbf{C3, readout under load.}
The cores of a circuit produce readout results at different rates, and the PS reads them with variable latency.
The readout must still drain while the program runs, and it must never stall the DSP.

%% file: sections/design.tex
\section{Technical Design}
\label{sec:design}

To address the three challenges of Section~\ref{sec:challenges}, \sys uses the DRAM of the control board as the main memory pool and BRAM as an explicitly managed staging buffer between DRAM and the DSP cores.
In this section, we first give an overview of this design (Section~\ref{sec:design-overview}).
Then, we describe the downlink (Section~\ref{sec:downlink}), which hides the latency of DRAM behind ping-pong buffers in BRAM (C1) and switches the buffers of all cores together at a constant cost while the compiler carries the oscillator phases across sub-circuits (C2).
Next, we describe the uplink (Section~\ref{sec:uplink}), which collects the readout of all cores in a shared buffer and drains it through a ring buffer in DRAM while the program runs (C3).
Finally, we describe the early-termination monitor (Section~\ref{sec:early-termination}), which uses the streamed readout to stop circuits that noise dominates.

\begin{figure}[t]
    \centering
    \includegraphics[width=\columnwidth]{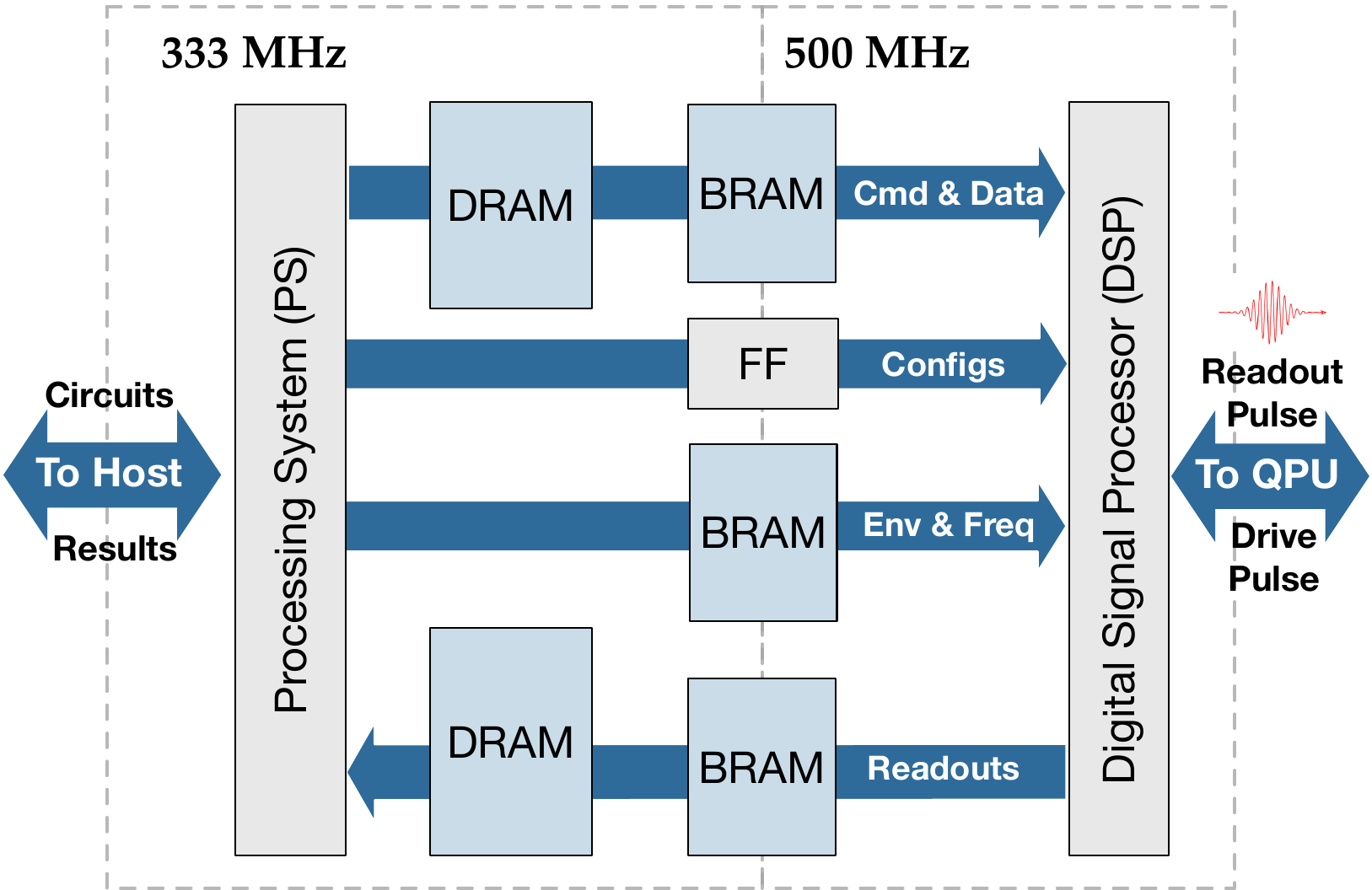}
    \caption{Design overview of \sys. Circuits flow from the processing system (PS) through the dynamic random-access memory (DRAM) and the block random-access memory (BRAM) of the programmable logic to the digital signal processing (DSP) cores. Readout results flow back through BRAM and DRAM to the PS.}
    \label{fig:overview}
\end{figure}

\subsection{Design Overview}
\label{sec:design-overview}
Figure~\ref{fig:overview} shows the design of \sys, which integrates DRAM with BRAM and flip-flops (FFs, which form the registers) on the PL to carry data in both the downlink and the uplink direction.
Table~\ref{tab:roles} lists where each part runs.
The host compiles the circuits, packs their commands and decodes the results.
Three servers on the PS move commands, configurations and results between the host and the PL, and the PL runs the circuits.

\input{tables/roles}

The downlink direction, from left to right in Fig.~\ref{fig:overview}, has three data paths.
The first data path carries circuit commands.
After the host compiles the circuits and sends their commands to the PS, the batch server on the PS moves them into DRAM on the PL by AXI DMA~\cite{amd_pg021}.
The PL then fetches the commands of a circuit into BRAM when the circuit is scheduled to run.
The second data path carries the configuration metadata of each circuit, such as the number of shots and the early-termination signal.
The PS writes these metadata directly into FFs in the PL over the AXI-Lite interface~\cite{arm_axi}, so the DSP cores read them from FFs without waiting.
The third data path carries the parameters that do not depend on the commands, such as envelope waveforms and frequencies.
The PS loads them into separate BRAM banks over AXI-Lite, also bypassing DRAM, because these parameters are shared across circuits and change only after calibration.
To keep them shared within a batch of different circuits, the compiler on the host compiles all circuits of a batch against one table layout, which places every envelope and frequency the batch needs at a fixed address.
The RPC server on the PS thus loads the tables once before the batch, and the circuits run back to back without reloading them.

The uplink direction, from right to left, has a single data path.
It offloads readout results from BRAM to DRAM and then to the PS by AXI DMA.
The readout results are accumulated in-phase and quadrature (IQ) data, each with a real and an imaginary part, and \sys annotates each result with a tag.
While the DSP cores keep driving the QPU for the remaining shots, the readout results of earlier shots already stream to the host.
A monitor on the PS or the host can analyze these results and decide whether noise has begun to dominate the computation.
If so, the stream server on the PS writes a stop signal into the FFs and terminates the DSP cores (Section~\ref{sec:early-termination}).
In our evaluation, this decision is preset.

DRAM and all its connections to the PS, BRAM and FFs run at a lower clock frequency, $333~\text{MHz}$ on the ZCU216, which is the user-interface clock of the DDR4 controllers~\cite{amd_pg150}.
BRAM, as the staging buffer, uses a dual-port configuration.
One port runs at the lower clock frequency, and the other runs at $500~\text{MHz}$ on the ZCU216 so that it adds no delay to the DSP.
Because the downlink and uplink data paths work at the same time, the execution and uplink of the current circuits overlap with the downlink of the next circuits in the schedule.

\begin{figure*}[t]
    \centering
    \includegraphics[width=\textwidth, trim=0 0 0 0, clip]{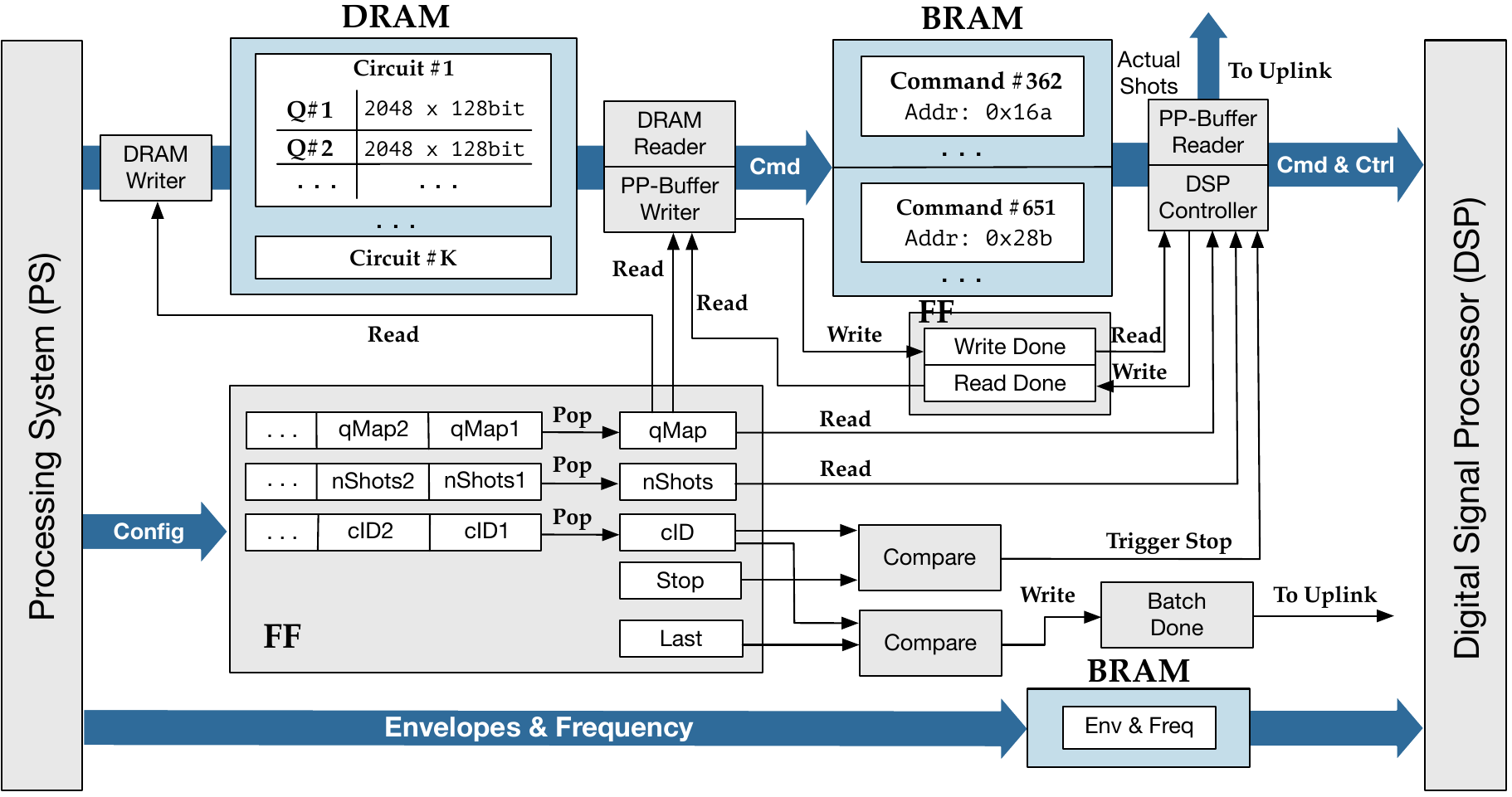}
    \caption{Design of the \sys downlink. The per-core commands of each circuit move from the dynamic random-access memory (DRAM) into one half of a ping-pong buffer in block random-access memory (BRAM) while the controller of the digital signal processing (DSP) cores executes the other half. The per-circuit metadata (\Code{qMap}, \Code{nShots}, \Code{cID}) are queued in flip-flops (FF).}
    \label{fig:downlink}
\end{figure*}

\subsection{\sys Downlink}
\label{sec:downlink}
The downlink of \sys has two goals.
The first goal is to execute deep experimental circuits whose size exceeds the total capacity of BRAM.
The second goal is to turn the punched-card workflow into a pipeline by executing circuits continuously.
For the first goal, \sys decomposes a deep circuit into a sequence of smaller sub-circuits, which is valid as long as no time interval, or only a pre-specified deterministic gap, separates them.
Both goals thus reduce to a single objective, which is to never stall the DSP cores and the QPU.

To meet this objective, the next circuit must already be in BRAM when the current circuit completes.
The downlink thus becomes a race between command consumption and data transfer.
Each circuit or sub-circuit must run at least as long as it takes to fetch the next one from DRAM, which we quantify as the refill condition in Section~\ref{sec:eval-rq1}.
The faster \sys transfers circuit commands, the more qubits and the denser circuits it supports.
\sys raises the effective transfer rate in two ways.
It overlaps circuit loading with execution, and it eliminates unnecessary padding.

To overlap loading with execution and hide the latency of DRAM (C1), \sys dedicates one BRAM bank to each DSP core and divides the bank into two address regions that form a ping-pong buffer (Fig.~\ref{fig:downlink}).
One region holds the commands of the current circuit, which the DSP core consumes.
The other region receives the commands of the next circuit from DRAM.
The read and write ports of the bank work independently, so execution and command loading proceed concurrently.
A pair of handshake registers, \Code{Write Done} and \Code{Read Done}, synchronizes the two halves.
The writer asserts \Code{Write Done} after filling a region, and the reader asserts \Code{Read Done} after draining one.
Each side polls the flag of the other before it accesses the opposite half.
The flags cross between the $333~\text{MHz}$ DRAM side and the $500~\text{MHz}$ DSP side through synchronizers (Appendix~\ref{sec:app-timing}).
To keep the lock-step cores together (C2), a command sequencer in the DSP clock domain starts all participating cores with one common start pulse once the next halves of all their buffers are ready.
It releases these halves together when the DSP reports the last shot, so all cores switch at the same boundary.

When the next halves are ready in time, the switch between the two halves of the ping-pong buffer takes a constant $134~\text{ns}$ (Section~\ref{sec:eval-rq1}).
At the end of the current circuit, the DSP core ends the program, waits until the last pulse leaves the signal pipeline and restarts from the other half.
Our cycle-accurate simulation of the command path around the \qubic DSP core~\cite{antq} shows that $120~\text{ns}$ of this handover is the restart of the DSP core itself, which the \qubic baseline also incurs between every two shots.
Only $14~\text{ns}$ comes from the handshake and bank switch of \sys (Appendix~\ref{sec:app-timing}).
Because the \qubic compiler starts the first pulse of each circuit $10~\text{ns}$ after the program begins, the next segment inside a circuit starts $144~\text{ns}$ later than in the unbroken program.

The \qubic compiler splits a deep circuit into sub-circuits after it has scheduled the whole circuit~\cite{qubic_distproc}.
It cuts at an idle point, a time at which no pulse or acquisition window is playing on any core.
Specifically, it takes the last idle point before a core would exceed the $2048$ commands of one buffer half.
Every operation keeps its start time in the unbroken schedule, counted from the cut.
A boundary at an idle point therefore delays everything after it by the same $144~\text{ns}$ on all cores and keeps their relative timing (Appendix~\ref{sec:app-timing}).
If no idle point lies within the capacity of a sub-circuit, the compiler cuts where the overlap is smallest and warns, naming the cores whose pulses are still playing and the length of the overlap.
On request, it refuses the split instead.
A cut can also be placed by hand, as in the boundary experiment of Section~\ref{sec:eval-device}, and the compiler checks it in the same way.
The compiler resolves virtual Z rotations~\cite{mckay2017virtualz} into the phases of the pulses and places a phase reset only at the start of the first sub-circuit, so the oscillator phases run on across every boundary, as C2 requires.
It splits only straight-line programs of pulses at fixed times and refuses programs with branches or run-time arithmetic (Section~\ref{sec:future_work}).
All sub-circuits share one table layout, so the envelopes and frequencies of the whole program must fit into the table memories, whose size \sys leaves unchanged.
RB reuses a few pulse shapes, so for RB this constraint does not grow with the sequence length.

To eliminate padding, \sys changes how circuits are laid out.
The punched-card workflow assumes that a circuit uses all DSP cores and pads the commands of unused cores with zeros, which wastes transfer bandwidth.
\sys instead derives metadata from the original circuit to address commands at a finer grain.
The \qubic client on the host packs the compiled commands of each circuit into an image that holds one command queue, the size of a buffer half, for each DSP core the circuit uses.
When the batch server moves circuits from the PS to DRAM through the DRAM Writer, they are therefore already organized as per-core pipelines laid out together (Fig.~\ref{fig:downlink}).
In this layout, the pipeline of an unused core is omitted.
The client records the used cores in a \Code{qMap} bitmap, which the batch server writes into the PL with the other metadata of the circuit.
The DRAM Reader \& PP-Buffer Writer module between DRAM and BRAM reads \Code{qMap} and fetches only the pipelines of enabled cores, which uses the full bandwidth of the DRAM--BRAM channel.
Each unused core loads a minimal idle program instead.
The PP-Buffer Reader \& DSP Controller module also uses \Code{qMap} to read and execute only the commands of enabled qubits.
\Code{qMap} is stored in FFs, apart from the per-core pipelines.
The \Code{qMap} entries of different circuits follow the same first-in, first-out (FIFO) order as the circuits in DRAM.
Because the batch server pushes the metadata of each circuit before its commands and the PL consumes both in circuit order, the two need no further synchronization.

More metadata are queued in FFs next to \Code{qMap}.
\Code{nShots} gives the number of shots of each circuit, and \Code{cID} gives its ID.
\sys matches the early-termination signal \Code{Stop} against \Code{cID} to make sure that it terminates the right circuit.
This check is necessary because of the latency of the feedback loop, which transfers readout data to the PS or the host, processes them in software and sends back the termination signal.
By the time the signal reaches the PL, the system may already have switched to a different circuit.
With early termination, a circuit may also stop before it executes all \Code{nShots}.
The DSP Controller therefore sends the number of shots each circuit actually ran, \Code{Actual Shots}, to the uplink, and the host uses it to decode the readout data.
Finally, \sys compares \Code{Last}, the \Code{cID} of the last circuit in the batch, with the current \Code{cID} to detect the end of the batch, which tells the \sys Uplink to drain all remaining data from the PL to the PS.

\begin{figure*}[t]
    \centering
    \includegraphics[width=1.0\textwidth]{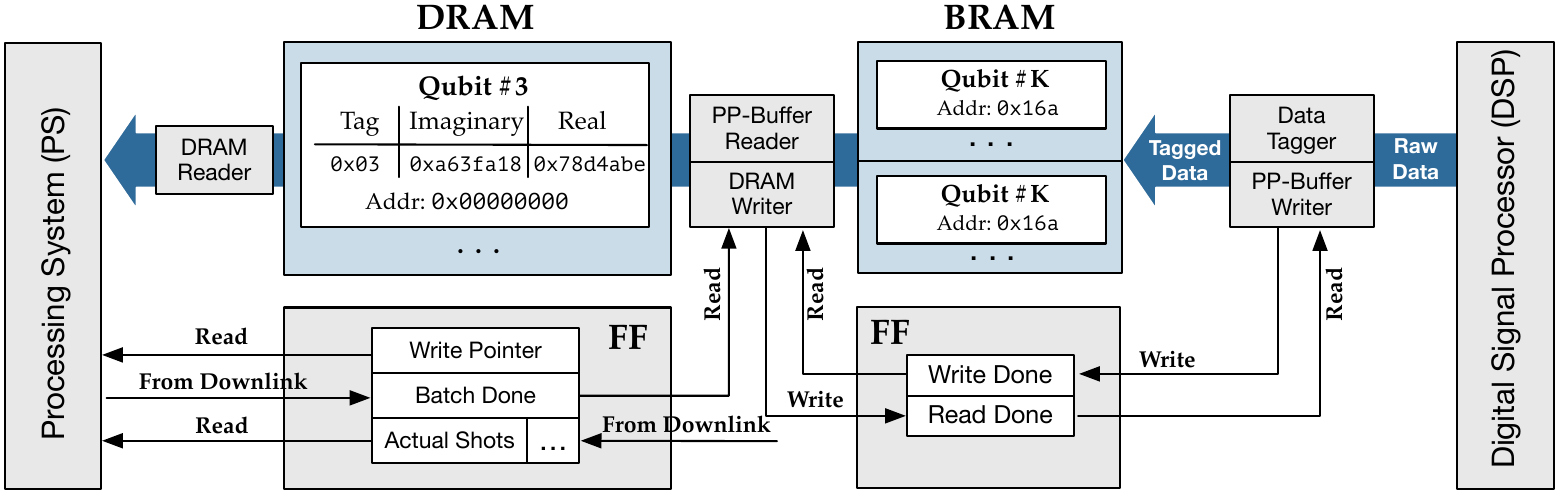}
    \caption{Design of the \sys uplink. Readout results of all digital signal processing (DSP) cores are tagged and collected in a shared ping-pong buffer, which is drained into a ring buffer in the dynamic random-access memory (DRAM) and read by the processing system (PS).}
    \label{fig:uplink}
\end{figure*}

\subsection{\sys Uplink}
\label{sec:uplink}
While the downlink streams circuits and the DSP cores execute them, the uplink must move the readout results to the PS and then to the host without stalling the DSP (C3).
This transfer prevents a backlog in long-duration, high-sampling-rate experiments and makes early termination possible.

\sys inserts DRAM as a middle layer between BRAM and the PS to absorb the uncertainty of the PS operating system and CPU.
Recall that DRAM has a lower worst-case latency than the PS and a large storage capacity, but its own access latency is still non-deterministic (Section~\ref{sec:challenges}).
In addition, each circuit has its own structure, so different DSP cores generate readout results at different rates.

To handle these uncertainties, \sys tags the readout results of all DSP cores and aggregates them into a shared ping-pong buffer of $64$ entries per half.
Each core places its latest result in a one-entry register, and a round-robin reader moves one entry from these registers into the buffer every $3$ cycles.
Two results of one core must therefore lie at least $64$ cycles ($128~\text{ns}$) apart, far less than a readout window, so the uplink never stalls the DSP.
At the highest readout rate of our benchmarks ($4.8~\text{MB/s}$), a half fills in $106.7~\mu\text{s}$, which is the time the DRAM Writer has to drain the other half.
A result lost to a longer stall would leave the readout stream shorter than expected, and the host rejects such a stream.
Every run reported here delivered the expected number of results.
Each entry carries an $8$-bit tag, a $28$-bit imaginary part and a $28$-bit real part (Fig.~\ref{fig:uplink}, DRAM region), and these widths can be configured for the channel count and the precision of the experiment.
The buffer reuses the ping-pong mechanism of Section~\ref{sec:downlink} with two differences.
It is shared across DSP cores instead of dedicated to one core.
It also switches halves either when the active half fills up or when a \Code{Batch Done} signal arrives from the \sys Downlink, after which the DRAM Writer drains the filled half into DRAM.

DRAM serves as a ring buffer.
The DRAM Writer maintains its Write Pointer, and the stream server on the PS maintains its Read Pointer, which is not stored in PL registers.
The large storage of DRAM absorbs the variable latency of both the PS and DRAM.
The stream server accepts a streamed batch only if its readout data fit into the $2~\text{GB}$ ring buffer, and the host requests the next batch only after it has received all readout data of the previous one.
Every batch therefore starts with an empty ring buffer, and the DRAM Writer never overwrites data that the PS has not read, however long the PS pauses.
At $4.8~\text{MB/s}$, the buffer holds about $7$ minutes of data.
Longer acquisitions would require the PL to see the Read Pointer, which it does not in the current design.
Besides readout data, the stream server reads metadata such as \Code{Actual Shots} from the PL and forwards them to the host, which needs them to decode the results.
Table~\ref{tab:contract} collects the conditions of the downlink and uplink and what happens when a condition is not met.

\input{tables/contract}

\subsection{Early-Termination Monitor}
\label{sec:early-termination}
With the \sys Uplink, the stream server on the PS receives the readout results of a circuit shortly after the circuit starts, and the host receives them soon after.
A monitor on the PS or the host can then measure how far noise has moved the measured distribution from the ideal one and decide whether the results are readable.
The monitor uses two metrics, the Total Variation Distance (TVD)~\cite{dahlhauser2021modeling} and the Hellinger Fidelity (HF)~\cite{hartnett2024learning}.

For two probability distributions $p$ and $q$ over measurement outcomes, the TVD is
\begin{equation}
  \mathrm{TVD}(p, q) = \frac{1}{2} \sum_{x} |p(x) - q(x)|,
\end{equation}
where $\mathrm{TVD} = 0$ indicates identical distributions and $\mathrm{TVD} = 1$ indicates disjoint supports.
The HF is
\begin{equation}
  \mathrm{HF}(p, q) = \left( \sum_{x} \sqrt{p(x) \cdot q(x)} \right)^2,
\end{equation}
where $\mathrm{HF} = 1$ indicates identical distributions and $\mathrm{HF} = 0$ indicates orthogonal distributions.
We call the results of a run \emph{unreadable} when $\mathrm{HF} \le 0.4$ and $\mathrm{TVD} \ge 0.6$, \ie, when its distribution has moved far from the ideal one.

The monitor evaluates the distribution of a circuit with $N$ shots at the checkpoints $k \lfloor N/10 \rfloor$ for $k = 1, \ldots, 9$ and at $N$ for $k = 10$, \ie, every $10\%$ of the shots, rounded down to whole shots.
Let $p_k$ denote the empirical distribution at checkpoint $k$ and $p_{\mathrm{ideal}}$ the ideal, noiseless output distribution of the circuit.
We regard the distribution as stable when both
\begin{equation}
  |\mathrm{TVD}(p_k, p_{\mathrm{ideal}}) - \mathrm{TVD}(p_{k-1}, p_{\mathrm{ideal}})| \le \epsilon_{\mathrm{TVD}}
\end{equation}
and
\begin{equation}
  |\mathrm{HF}(p_k, p_{\mathrm{ideal}}) - \mathrm{HF}(p_{k-1}, p_{\mathrm{ideal}})| \le \epsilon_{\mathrm{HF}}
\end{equation}
hold, where $\epsilon_{\mathrm{TVD}} = \epsilon_{\mathrm{HF}} = 0.03$.

The monitor sends an early-termination signal to the PL when the distribution is stable under both metrics and the results are unreadable, but not before the third checkpoint ($30\%$ of the shots).
This lower bound avoids premature termination on the statistical fluctuations of the first shots.
Because consecutive checkpoints share most of their shots, a small change between them does not by itself show that the estimate has converged.
Section~\ref{sec:eval-stop} therefore measures how often the rule stops runs that are not unreadable.

In our evaluation, the monitor does not yet run as a live service (Section~\ref{sec:future_work}).
Instead, the host sends the stream server the stop checkpoint of each circuit, chosen within the range given by a noise simulation study (Section~\ref{sec:eval-stop}).
At every checkpoint, the stream server waits for the time one evaluation takes on the ARM cores of the PS, and it writes \Code{Stop} into the PL when it reaches the preset checkpoint.
The rule compares each distribution with the ideal one, so it applies only to circuits whose ideal output distribution is known.

%% file: tables/roles.tex
\begin{table*}[t]
\centering
\caption{Where the parts of \sys run. The host is the user's computer, the PS is the ARM processor of the ZCU216 running Linux, and the PL is the FPGA fabric.}
\label{tab:roles}
\setlength{\aboverulesep}{0pt}
\setlength{\belowrulesep}{0pt}
\renewcommand{\arraystretch}{1.3}
\footnotesize
\begin{tabularx}{\textwidth}{>{\raggedright\arraybackslash}p{0.07\textwidth}|>{\raggedright\arraybackslash}p{0.2\textwidth}|>{\raggedright\arraybackslash}X}
\toprule
\rule{0pt}{2.5ex}Where & Part & Role \\
\midrule
Host & \qubic compiler~\cite{qubic_distproc} & Compiles and assembles each circuit into per-core commands and envelope and frequency tables. Splits a deep circuit into sub-circuits at idle points (Section~\ref{sec:downlink}). With the shared table layout of the assembler, all circuits of a batch use one set of tables. \\
 & \qubic client~\cite{qubic_software} & Groups circuits with identical tables into hardware batches. Packs the commands of each circuit into an image without the unused cores and records the used cores in \Code{qMap}. Sends the tables, the batches with their images and the stop table to the PS. Receives the readout stream and splits it into the results of each circuit. \\
\midrule
PS & RPC server~\cite{qubic_software} & Loads the bitfile. Before a batch, resets the DSP cores and writes the envelope and frequency tables into BRAM. \\
 & Batch server~\cite{qubic_software} & Holds the command images in a $64~\text{MiB}$ circuit buffer, moves them into DRAM by DMA, writes \Code{nShots} and \Code{qMap} of each circuit and \Code{Last} into the PL, and starts the batch. \\
 & Stream server~\cite{qubic_software} & Reads the readout data and \Code{Actual Shots} from the PL as they arrive, keeps the Read Pointer of the DRAM ring buffer and forwards the data to the host. Applies the stop table and writes \Code{Stop} into the PL. \\
\midrule
PL & \sys gateware~\cite{qubic_gateware} & Fetches the commands from DRAM into the ping-pong buffers, switches the buffers of all cores together, collects and tags the readout results and writes them into the DRAM ring buffer, and flags boundary waits above $16~\text{ns}$. \\
\bottomrule
\end{tabularx}
\end{table*}

%% file: tables/contract.tex
\begin{table*}[t]
\centering
\caption{Operating conditions of \sys, with what each part of a run requires, what the system checks, and what happens when the requirement is not met.}
\label{tab:contract}
\setlength{\aboverulesep}{0pt}
\setlength{\belowrulesep}{0pt}
\renewcommand{\arraystretch}{1.3}
\footnotesize
\begin{tabularx}{\textwidth}{>{\raggedright\arraybackslash}p{0.15\textwidth}|>{\raggedright\arraybackslash}X|>{\raggedright\arraybackslash}X|>{\raggedright\arraybackslash}X}
\toprule
\rule{0pt}{2.5ex}Part & Requirement & Check & When the requirement is not met \\
\midrule
Command refill & Each circuit lasts at least $t_0 + n' S/B_D$, the time to fetch the next one (Section~\ref{sec:eval-rq1}) & A sticky flag for any boundary wait above $16~\text{ns}$ and the wait at the last boundary, returned with the results & The DSP waits for the commands and continues. The wait adds idle time at that boundary, and the run is not rejected. \\
Circuit splitting & A straight-line program, every cut at an idle point, at most $2048$ commands per core in a sub-circuit, the tables of the whole program fit (Section~\ref{sec:downlink}) & Before the run, the compiler checks the program structure, the capacity of every sub-circuit and whether each cut lies at an idle point, and the assembler checks the table memories & A program with branches or run-time arithmetic, an over-full sub-circuit or tables that do not fit are refused. At a cut that is not idle, the compiler warns (on request refuses) and reports the overlap, by which everything after the cut is delayed beyond the $144~\text{ns}$. \\
Readout uplink & Two results of one core at least $64$ cycles apart, and a half drained before the other fills (Section~\ref{sec:uplink}) & The number of results delivered against the number expected & A lost result leaves the stream short, and the host rejects the run. \\
DRAM ring buffer & The readout data of a batch fit into $2~\text{GB}$, with one batch at a time & Admission check before the batch & The PS refuses the batch. \\
Host link & The TCP stream from the PS to the host stays open & Length of the stream & The run fails, and the measurement script restarts the services and repeats it. \\
\bottomrule
\end{tabularx}
\end{table*}

%% file: sections/implementation.tex
\section{Implementation}
\label{sec:implementation}
All experiments run on AMD Xilinx ZCU216 RFSoC evaluation boards~\cite{amd_ug1390}, whose Zynq UltraScale+ device~\cite{amd_ds889} combines a quad-core ARM Cortex-A53 (the PS) with programmable logic (the PL).
The DSP cores run at $500~\text{MHz}$, and the PL has $4~\text{GB}$ of DRAM, split between two DDR4 controllers.
The ZCU216 is the standard deployment platform of \qubic and a widely adopted RFSoC board in the community, so our results transfer to existing deployments without a board redesign.
Both the \sys Downlink and Uplink have been merged into the \texttt{feat/ddr\_mem} branch of the \textsl{QubiC} gateware and software repositories~\cite{qubic_gateware, qubic_software}.
The versions used in our evaluation, gateware commit \texttt{38543b7} and software commit \texttt{29d123a}, are pinned in~\cite{antq}.
The shared table layout of a batch is an option of the \qubic assembler, which has been merged into its \texttt{distributed\_processor} package~\cite{qubic_distproc}.
The splitting of deep circuits is a pass of the \qubic compiler in the same package and has been submitted for merging into its \texttt{feat/ddr\_mem} branch.
The version used in our evaluation, commit \texttt{c2adc1c}, is pinned in~\cite{antq}.

We build three bitfiles (FPGA configuration files) for our evaluation.
Each is named by the commit it was built from, and all three meet the timing constraints of the $500~\text{MHz}$ clock.
For the bench experiments, we use the $14$-core \sys bitfile (\texttt{13b440c3}) and a $14$-core bitfile with the \qubic Downlink and the \sys Uplink (\texttt{0071783e}), which also contains the original BRAM paths used by the \qubic baseline.
For the experiments on the real QPU, we use the $8$-core \sys bitfile (\texttt{c26942c0}), and the baseline is the stock \qubic bitfile that the testbed uses in daily operation.

Both \qubic and \sys use the configuration with $14$ drive and $14$ readout channels, which controls $14$ qubits, the maximum that a single board supports.
For the \sys Downlink, DRAM is $2~\text{GB}$, and each of the $14$ ping-pong buffers uses $64~\text{KiB}$ of BRAM.
\Code{qMap} is $16$-bit (one bit per DSP core), \Code{nShots} is $32$-bit, and \Code{cID} is $16$-bit.
Command data are $128$-bit, and each half of a ping-pong buffer holds $2048$ commands.
The ring buffer in DRAM holds up to $1024$ circuits.
For the \sys Uplink, DRAM is also $2~\text{GB}$.
Raw readout data are $64$-bit, with a $32$-bit imaginary and a $32$-bit real part.
The tagged data keep an $8$-bit tag and the upper $28$ bits of each component.
Dropping the $4$ least significant bits changes a value by at most $16$, while one readout blob on our qubit spreads over about $3.4 \times 10^5$ in the same units.
Each half of the uplink ping-pong buffer holds $64$ readout entries.
The \Code{Write Pointer} and \Code{Actual Shots} are both $32$-bit values.

\input{tables/resources}

Table~\ref{tab:resource_utilization} summarizes the FPGA resources of the three designs.
The full \sys raises the LUT utilization from $22.14\%$ to $32.35\%$ and the FF utilization from $22.48\%$ to $29.93\%$, mostly for the control logic of the AXI DMA transfers, the ring-buffer pointers and the circuit configurations.
The \sys Uplink alone accounts for about half of this increase.
The BRAM utilization rises from $57.41\%$ to $71.94\%$ for the ping-pong buffers that absorb the non-deterministic DRAM latency.
The DSP utilization stays almost constant, since the design changes the control and memory layers and leaves the pulse generation untouched.
The \qubic baseline uses no DRAM, whereas \sys uses all $4~\text{GB}$ on the board.
The full design fits within the resources of the ZCU216.

%% file: tables/resources.tex
\begin{table}[t]
\centering
\caption{Resource utilization of the placed designs in the $14$-core configuration. The table gives the look-up tables (LUTs), flip-flops (FFs), $36$-kbit block random-access memories (BRAM36) and digital signal processing (DSP) slices as a percentage of those of the ZCU216 field-programmable gate array, and the external dynamic random-access memory (DRAM) used.}
\label{tab:resource_utilization}
\setlength{\aboverulesep}{0pt}
\setlength{\belowrulesep}{0pt}
\renewcommand{\arraystretch}{1.4}
\setlength{\tabcolsep}{5pt}
\scriptsize
\begin{tabular}{l|c|c|c}
\toprule
\rule{0pt}{3ex}Resource
& \makecell{\qubic\\Downlink\\+ Uplink}
& \makecell{\qubic\\Downlink +\\\sys Uplink}
& \makecell{\sys\\Downlink\\+ Uplink} \\
\midrule
LUTs    & $22.14\%$ & $27.32\%$ & $32.35\%$ \\
FFs     & $22.48\%$ & $26.09\%$ & $29.93\%$ \\
BRAM36  & $57.41\%$ & $61.02\%$ & $71.94\%$ \\
DSPs    & $67.18\%$ & $67.25\%$ & $67.32\%$ \\
DRAM    & $0~\text{GB}$ & $2~\text{GB}$ & $4~\text{GB}$ \\
\bottomrule
\end{tabular}
\end{table}

%% file: sections/evaluation.tex
\section{Evaluation}
\label{sec:eval}
We evaluate \sys on three questions.
Q1 asks whether \sys runs experiments that exceed the memory capacity of \qubic, on the bench (Section~\ref{sec:eval-rq1}) and on a superconducting qubit (Sections~\ref{sec:eval-device} and~\ref{sec:eval-noise}).
Q2 asks how much \sys raises the throughput of computing tasks (Section~\ref{sec:eval-rq2}).
Q3 asks how \sys supports the early termination of noise-dominated circuits (Section~\ref{sec:eval-stop}).
We compare three configurations (Table~\ref{tab:capabilities}), namely the \qubic baseline (\qubic Downlink + Uplink)~\cite{xu2023qubic}, the \qubic Downlink with the \sys Uplink (uplink-only) and the full \sys (\sys Downlink + Uplink).
Appendices~\ref{sec:app-timing} to~\ref{sec:app-stop} give the details of the measurements and analyses.

\input{tables/functionality}

\subsection{Experiments Beyond BRAM Capacity}
\label{sec:eval-rq1}
To answer Q1 on the bench, we reconstruct four benchmark experiments from the literature that exceed the memory of \qubic (Table~\ref{tab:capabilities}).
Experiment (1) is a repeated Ramsey sequence for $1/f$ noise spectroscopy with $50000$ shots and an idle interval of $2~\text{ms}$ after each readout, \ie, one shot every $2003.648~\mu\text{s}$~\cite{yan2012noise}.
Experiment (2) is a Ramsey-type charge-parity detector with $8000$ samples per trace and an idle interval of $6~\mu\text{s}$, \ie, one shot every $8.932~\mu\text{s}$~\cite{riste2013charge}.
Experiment (3) is single-qubit \emph{restless} RB, without qubit reset between shots, at $5101$ Cliffords~\cite{tornow2022restless}.
Experiment (4) is two-qubit subspace leakage error randomized benchmarking (SLERB) at $500$ Cliffords~\cite{hughes2025trappedion}.
Experiments (1) and (2) require long readout records with precise inter-shot timing, and (3) and (4) require deep sequences of random Cliffords.
We translate all four into pulse-level commands for both stacks, about $2$ commands per qubit for a one-qubit (1Q) Clifford and about $10$ for a two-qubit (2Q) Clifford.
The commands use the default qubit configuration of \qubic, with fast reset and readout of $2.83$ and $2.60~\mu\text{s}$ and 1Q and 2Q gates of $24$ and $332~\text{ns}$.
The programs and the configuration are released in~\cite{antq}.

The \qubic baseline runs none of the four.
Experiments (1) and (2) need more than the $1024$ readout records per qubit that BRAM holds, and their fixed shot spacing rules out splitting them into separate runs.
The \sys Uplink supports both, because it streams the readout results to the host during the run.
Experiments (3) and (4) exceed the command buffer of $2048$ commands, which holds about $1000$ Cliffords, and their random Cliffords cannot be rolled into loops.
With the \sys Downlink, they run as streams of sub-circuits that share one table layout (Section~\ref{sec:downlink}).

\begin{figure*}[t]
    \centering
    \includegraphics[width=\textwidth]{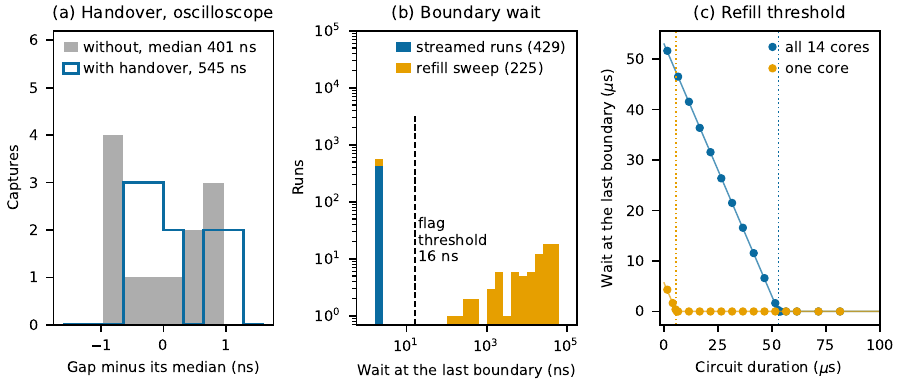}
    \caption{Timing on the $14$-core \sys bitfile. (a) Gap between two marker pulses on the oscilloscope with and without a command-buffer handover between them ($12$ captures each), relative to the median of each set. (b) Wait at the last sub-circuit boundary of every streamed run on the bench and on the qubit and of every run of the refill sweep. The gateware flags waits above $16~\text{ns}$ (dashed). (c) The refill sweep, showing that wait versus the circuit duration for all $14$ cores and for one core, with the thresholds $t_0 + nS/B_D$ (dotted) and the shortfall they predict (lines).}
    \label{fig:timing}
\end{figure*}

We run all four experiments on the $14$-core \sys bitfile (Fig.~\ref{fig:timing}).
An oscilloscope shows that (1) and (2) execute every shot at the programmed period, over $10$ records of about $500$ shots for (1) and $3$ records of $8000$ shots for (2).
When the next command buffers are ready in time, a handover follows a fixed sequence of $67$ DSP cycles ($134~\text{ns}$) in our cycle-accurate simulation (Appendix~\ref{sec:app-timing}).
On the oscilloscope, a handover between two marker pulses lengthens their gap from $401$ to $545~\text{ns}$ with an unchanged spread, and $10~\text{ns}$ of the difference is the first-pulse offset of the next segment (Fig.~\ref{fig:timing}a).
The gateware records the wait at the last boundary of a run and flags a run if any of its boundaries waits more than $16~\text{ns}$ for its commands.
It does not record the waits of the other boundaries.
In every streamed run, on the bench and on the qubit, the flag is clear and the last boundary waits the nominal one cycle ($2~\text{ns}$, Fig.~\ref{fig:timing}b).
The last handover thus follows the simulated sequence exactly, and every other handover takes at most $14~\text{ns}$ longer.
These runs include batches that stream commands and readout at the same time on $1$ to $14$ cores.
Experiments (3) and (4) have $5119$ and $14999$ boundaries per run, and their run times exceed the programmed durations by $0.34\%$ and $0.11\%$, mostly the cost of their handovers.

A handover takes this constant time only if the next commands are ready.
Each circuit or sub-circuit must therefore run at least as long as it takes to bring the next one into BRAM.
Let $n$ be the number of DSP cores a circuit uses, $S = 32~\text{KiB}$ the command data of one core (one half of a ping-pong buffer, \ie, $2048$ commands of $128$ bits), $K$ the number of circuits in a batch, $N_R = 1024$ the circuits held by the ring buffer in DRAM, and $N_P$ the circuits held by the circuit buffer on the PS ($146$ for the $14$-core bitfile).
With both buffers filled before the start, a batch of circuits of duration $T_c$ each runs without a stall only if
\begin{align}
T_c \ge \max\big\{\, & t_0 + nS/B_D,\ \left(1 - N_R/K\right)^{+} (t_0 + nS/B_P), \nonumber\\
 & \left(1 - (N_R + N_P)/K\right)^{+} nS/B_H \,\big\},
\label{eq:refill}
\end{align}
where $(x)^{+} = \max(x, 0)$, $t_0$ is a fixed latency, and $B_D$, $B_P$ and $B_H$ are the transfer rates from DRAM to BRAM, from the PS to DRAM and from the host to the PS.
The first term is the time to fetch the next circuit from DRAM and applies to every transition.
The other two terms are amortized bounds for batches larger than what DRAM and the PS hold.
Condition~(\ref{eq:refill}) is necessary under these assumptions, and we fit its parameters to our measurements (Appendix~\ref{sec:app-timing}).
A sweep of the circuit duration gives $B_D \approx 9.0~\text{GB/s}$ and $t_0 \approx 2.0~\mu\text{s}$, \ie, thresholds of $5.6~\mu\text{s}$ for one core and $53.0~\mu\text{s}$ for $14$ cores (Fig.~\ref{fig:timing}c).
Below its threshold, the DSP waits for the missing commands and then continues, and the flag records the wait.
For batches much larger than what DRAM and the PS hold, the other two terms reach $8.3$ and $90.4~\mu\text{s}$ for one and $14$ cores when the circuits come from the PS ($B_P = 5.19~\text{GB/s}$), and $0.27$ and $3.82~\text{ms}$ when they come from the host ($B_H \approx 0.12~\text{GB/s}$).
All circuits of our evaluation stay above their thresholds.
The shortest sub-circuits, those of (3) and (4), last $49.1$ and $81.4~\mu\text{s}$ on $1$ and $2$ cores.
Before a run, condition~(\ref{eq:refill}) with the fitted parameters indicates whether the refills can keep up.
After the run, the flag and the recorded wait of the last boundary show whether they did.
Table~\ref{tab:contract} summarizes these operating conditions.

\begin{figure*}[t]
    \centering
    \includegraphics[width=\textwidth]{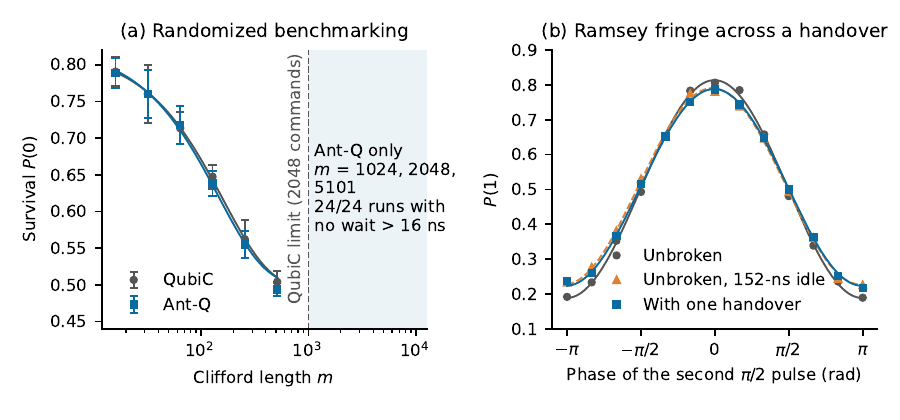}
    \caption{Validation on a superconducting qubit. (a) Survival probability $P(0)$, \ie, the probability of measuring the qubit back in its initial state, of single-qubit RB versus the sequence length $m$ that both control stacks can hold. Points are the mean over $8$ random sequences of $2048$ shots with the pointwise $95\%$ confidence interval of that mean. Curves are fits $B + A p^m$~\cite{magesan2011scalable} whose floor $B$ comes from the readout calibration. \sys additionally executes $1024$, $2048$ and $5101$ Cliffords, which exceed the command buffer of \qubic (dashed). (b) Ramsey fringe, the probability $P(1)$ of measuring the excited state versus the phase of the second $\pi/2$ pulse, with one command-buffer handover inside the free evolution, compared with the unbroken program and with the unbroken program that idles for $152~\text{ns}$ at the place of the handover.}
    \label{fig:device}
\end{figure*}

\subsection{Validation on a Superconducting Qubit}
\label{sec:eval-device}
The bench shows that these experiments execute with the expected timing.
To answer Q1 on a qubit, we validate \sys on one transmon qubit ($T_1 = 48~\mu\text{s}$, $T_2^* = 1.4~\mu\text{s}$) of an $8$-qubit fixed-frequency device at the Advanced Quantum Testbed~\cite{aqt}.
All measurements are taken in one session with the same calibrated pulses, using our $8$-core \sys bitfile and the stock \qubic bitfile of the testbed, each with its own software.
We refer to them as the \sys and \qubic stacks.

We first compare the two stacks in four alternating blocks (\qubic, \sys, \qubic, \sys) on seven pre-registered endpoints~\cite{antq} (Table~\ref{tab:device_endpoints}).
The endpoints are the readout assignment fidelity, the phase and contrast of a Ramsey fringe, a timing loop that repeats a Ramsey sequence every $6~\mu\text{s}$ for $8000$ shots (the device version of (2)), the mean and variance of paired Ramsey differences (Section~\ref{sec:eval-noise}), and the error per Clifford (EPC) of 1Q RB.
An endpoint counts as equivalent when the confidence interval (CI) of the difference between the stacks lies within a pre-registered margin in both passes and each stack stays within the margin between its own two blocks.
Six endpoints meet this rule.
The assignment fidelity does not, because the baseline changed between its own two blocks by as much as the two stacks differ.
This comparison therefore cannot separate a difference between the stacks from drift.
For the timing loop, \sys runs all $8000$ shots as one program at the compiled spacing of $6.000~\mu\text{s}$, as the oscilloscope shows for (2) on the bench.
\qubic, whose readout buffer holds $1024$ records, splits the loop into runs of at most $1024$ shots, and the host round trips between the runs break the fixed spacing.
The EPC of \sys differs from that of \qubic by $-0.0003$ and $-0.0008$ in the two passes, at an EPC of $0.0024$ to $0.0034$.
Because the pre-registered margin of $\pm 0.005$ is wider than the EPC itself, we bound this difference more tightly over the RB depths below.
The acquisition departs from the plan in its block order and omits the planned transport controls (Appendix~\ref{sec:app-device}).

\input{tables/device_endpoints}

Next, we place one command-buffer handover inside the free evolution of the Ramsey fringe and in the middle of 1Q RB sequences.
We compare these programs with the same programs unbroken and with unbroken programs that idle for $152~\text{ns}$ at the same place (Fig.~\ref{fig:device}b).
Against the idling programs, the handover changes neither the fringe phase ($+0.038~\text{rad}$, 95\% CI $[-0.004, 0.082]$) nor its contrast (ratio $1.002$, 95\% CI $[0.967, 1.040]$).
Against the unbroken programs, the contrast falls to $0.903$ of its value, as an exponential $T_2^*$ decay predicts for the $144~\text{ns}$ that the handover adds.
A handover inside a circuit thus acts on the qubit as an idle of $144~\text{ns}$.
In long RB sequences, a handover recurs about once per $1020$ Cliffords.
Extrapolated from this one boundary under the Markovian noise model of the RB fit~\cite{magesan2011scalable}, it adds at most $3 \times 10^{-5}$ to the error per Clifford (Appendix~\ref{sec:app-handover}).

Finally, we run 1Q RB up to $5101$ Cliffords with $8$ random sequences of $2048$ shots per length (Fig.~\ref{fig:device}a).
The \qubic baseline rejects $1024$ Cliffords, which compile to $2055$ commands.
\sys executes $1024$, $2048$ and $5101$ Cliffords as streams of $2$, $3$ and $5$ sub-circuits, with no boundary wait above $16~\text{ns}$ in all $24$ runs.
The $5101$-Clifford sequence runs its $2048$ shots in a median of $1804.08~\text{ms}$ against the programmed $1802.50~\text{ms}$, and its handovers account for $1.47~\text{ms}$ of the difference.
Where both stacks run ($m \le 512$), the EPC of \sys differs from that of \qubic by $+0.0003$ (95\% CI $[-0.0001, 0.0007]$).
This interval allows an increase of at most $0.0007$, less than a quarter of the EPC ($0.0031$ and $0.0034$).
From $1024$ Cliffords on, the decay on this qubit has saturated ($A p^m = 3 \times 10^{-4}$ at $m = 1024$).
Results at these depths therefore demonstrate executability but do not provide meaningful fidelity estimates.
Their survival data are released in~\cite{antq}.

\subsection{Noise Spectroscopy on the Qubit}
\label{sec:eval-noise}
\begin{figure*}[!t]
    \centering
    \includegraphics[width=0.62\textwidth]{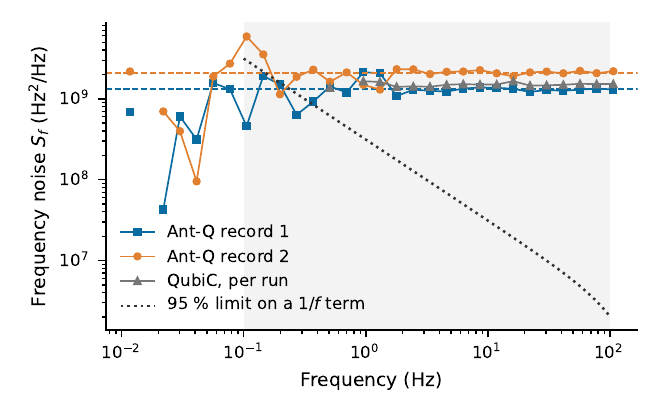}
    \caption{Spectrum of the pair differences, converted to frequency noise with the calibrated slope, from the two uninterrupted $100~\text{s}$ paired Ramsey records on \sys, with the binomial floor of each record (dashed), the \qubic baseline analysed run by run (mean of its two records, from $0.49~\text{Hz}$), the analysis band (shaded) and the nominal $95\%$ upper limit on a $1/f$ component (dotted). Each point is the periodogram averaged in one of $30$ logarithmic frequency bins.}
    \label{fig:noise}
\end{figure*}
The paired Ramsey records of the endpoint comparison also serve as noise spectroscopy~\cite{yan2012noise}.
Each paired Ramsey record holds $25000$ pairs of Ramsey measurements at opposite quadrature phases, with one measurement every $2~\text{ms}$.
The outcome difference of a pair follows the qubit frequency, while a drift of the readout cancels.
\sys acquires a record as one uninterrupted $100~\text{s}$ run, so its spectrum has a frequency spacing of $0.01~\text{Hz}$.
The \qubic baseline acquires the same pairs in runs of $2.048~\text{s}$, separated by gaps that its software does not timestamp.
We therefore analyse its records run by run, and their spectra start at $0.49~\text{Hz}$.
Both \sys records are consistent with white noise at the binomial floor of the single-shot outcomes from $0.1$ to $100~\text{Hz}$, and no $1/f$ component is resolved (Fig.~\ref{fig:noise}).
Under a Gaussian $1/f$ noise model (Appendix~\ref{sec:app-noise}), the nominal $95\%$ upper limit on the $1/f$ level at $1~\text{Hz}$ is $17.7~\text{kHz}/\sqrt{\text{Hz}}$, against $25.3~\text{kHz}/\sqrt{\text{Hz}}$ for the same records analysed in the runs of the baseline.
In simulated records, the whole records detect a $1/f$ component, with $50\%$ probability at a $5\%$ false-alarm rate, at spectral densities $1.37$ and $1.62$ times lower than the run-by-run analysis of the same records.
Appendix~\ref{sec:app-noise} gives the analysis, its departures from the pre-registered plan and its sensitivity to the model.

Table~\ref{tab:physics} collects what the four experiments gain from \sys on the bench and on the qubit.
Three of these gains are measured directly.
They are the uninterrupted acquisition with fixed shot spacing, the lower frequencies that this acquisition reaches, and the depth of the executable RB sequences ($5101$ against about $1000$ Cliffords).
The uninterrupted acquisition keeps all $8000$ shots of the timing loop of (2) at the compiled spacing and takes (1) as one $100~\text{s}$ record instead of $2.048~\text{s}$ runs.
Two further gains also depend on assumptions beyond the measurement.
The tighter limit on $1/f$ noise rests on the noise model, and the higher sensitivity rests on simulated records.

\input{tables/physics}

\subsection{Computing Throughput Improvement}
\label{sec:eval-rq2}
To answer Q2, we build a test set of $20$ quantum computing circuits (Table~\ref{tab:speedup}) from prior works~\cite{baumer2021scalable, liang2024aishot, teegarden2026threemonths, belaloui2025vqe, amico2019shor, hao2024qaoa, gharibyan2025qml, chawla2024srh, vemula2022grover, agnihotri2026qsvm} and from tutorials of IBM, Rigetti and Open Quantum Design~\cite{ibm_teleportation, ibm_query_algorithms, grove_vqe, oqd_qaoa}.
We exclude circuits with more than $14$ qubits, which one ZCU216 board cannot drive and read out.
The set covers Greenberger--Horne--Zeilinger (GHZ) states~\cite{greenberger1989ghz}, the variational quantum eigensolver (VQE)~\cite{peruzzo2014vqe}, the quantum approximate optimization algorithm (QAOA)~\cite{farhi2014qaoa}, Grover search~\cite{grover1996search}, Shor factoring~\cite{shor1997factoring}, the quantum Fourier transform (QFT)~\cite{coppersmith1994qft} and quantum machine learning (QML) with a quantum support vector machine (QSVM)~\cite{havlicek2019qsvm}.
In Table~\ref{tab:speedup}, PDM in VQE SrH PDM denotes the permanent dipole moment.
We construct the circuits that are not available and translate all of them for \qubic~\cite{antq}.
For QAOA MaxCut and VQE HeH$^+$, whose sources state a shot budget per problem instance and per energy evaluation, we use that budget as the shot count.
Our circuits read each qubit twice per shot, once for fast reset and once for the final measurement.
The $1024$ readout records per qubit of the baseline therefore hold $512$ shots, and the $16$ circuits with more than $512$ shots run on the baseline in sequential runs of at most $512$ shots.
We also build $10$ random batches of $30$ circuits and two homogeneous batches of short (GHZ-8) and deep (QAOA MaxCut) circuits.

\input{tables/speedup}

We time each circuit or batch on the PS, from the first command written to BRAM (\qubic Downlink) or to DRAM (\sys Downlink) to the last readout data in PS memory.
All circuits are staged in PS memory beforehand, so the compilation, the transfer from the host and the scheduling lie outside this interval.
The PS procedure of the baseline is implemented in C.
Each circuit or batch is measured five times after one warm-up run, and we report the mean.
The reference time is the number of shots times the per-shot time measured on the baseline bitfile.
The excess run time $\Delta$ of a configuration is its execution time minus this reference (Appendix~\ref{sec:app-throughput}).

On the \qubic baseline, $\Delta$ is $18.91\%$ to $313.04\%$ of the reference time (Table~\ref{tab:speedup}) and exceeds the reference itself for three circuits.
On a connected device, the QPU would idle during this time.
The \sys Uplink alone reduces $\Delta$ to $0.74\%$--$274.41\%$ by overlapping the readout transfer with the execution.
The full \sys also overlaps the circuit transfer with the execution, omits the commands of unused cores and moves circuits by DMA through DRAM.
Its $\Delta$ is at most $0.53\%$ for circuits with at least $1000$ shots.
The remainder, at most $0.23~\text{ms}$ per circuit, has two parts.
One is the fixed cost of uploading, starting and draining a circuit ($38.3$ to $220.9~\mu\text{s}$).
The other is the per-shot difference between the two bitfiles, which makes $\Delta$ slightly negative for three circuits (Appendix~\ref{sec:app-throughput}).
Smaller and shorter circuits gain the most, since the reference time dominates long circuits.

\input{tables/speedup_batch}

Batches show the same trend.
For the $30$-circuit batches (Table~\ref{tab:batch_speedup}), $\Delta$ falls from $312.49\%$ to $0.87\%$ for the batch of short circuits and from $4772.55~\text{ms}$ to $3.10~\text{ms}$ for the batch of deep circuits.
Over the ten random batches, the throughput gain $t_{\qubic} / t_{\sys} - 1$, with $t$ the execution time, averages $51.06\%$ (95\% CI $46.85\%$--$55.28\%$) against the baseline compiled with the same shared table layout as \sys.
Our measurement plan designated this comparison as confirmatory.
Against the stock baseline, which reloads the tables of every circuit, the gain averages $59.90\%$ (95\% CI $55.38\%$--$64.42\%$).
These numbers are means of the gains of the individual batches, and the ratios of the mean times in Table~\ref{tab:batch_speedup} give $50.52\%$ and $59.08\%$.
The \sys Uplink alone gains $40.96\%$ against the stock baseline, so most of the gain on the random batches comes from the uplink.
With shared tables on both sides, adding the \sys Downlink to the \sys Uplink gains $4.11\%$ (95\% CI $3.74\%$--$4.49\%$) on the random batches, $286.00\%$ on the batch of short circuits and $1.34\%$ on the batch of deep circuits.
The downlink thus matters most for short circuits.

\input{tables/noise_analysis}

\subsection{Application: Early Termination}
\label{sec:eval-stop}
To answer Q3, we first find the circuits that noise drives to unreadable results in simulation.
We simulate every circuit of Section~\ref{sec:eval-rq2} with its shot count of Table~\ref{tab:speedup} and $20$ simulator seeds on six noise simulation backends of Qiskit~\cite{Qiskit} (Qiskit 2.3.1, Qiskit Aer 0.17.2, Qiskit IBM Runtime 0.46.1).
The backends, FakeManilaV2, FakeLagosV2, FakeGuadalupeV2, FakeAlgiers, FakeSherbrooke and FakeTorino ($5$ to $133$ qubits), are calibration snapshots of IBM Quantum devices with their coherence times, gate errors and readout-assignment errors.
The sources of VQE BeH$_2$ and QML Image Class publish no circuit parameters, so we use one fixed synthetic parameter set for each.
QAOA MaxCut uses the optimal angles of its single-layer ($p = 1$) circuit~\cite{antq}.
Against the exact noiseless distribution, a run is graded \textsc{Pass} if $\mathrm{HF} > 0.7$ and $\mathrm{TVD} < 0.3$, \textsc{Fail} (unreadable) if $\mathrm{HF} \le 0.4$ and $\mathrm{TVD} \ge 0.6$, and \textsc{Marginal} otherwise~\cite{hartnett2024learning, dahlhauser2021modeling}.
The \textsc{Marginal} band avoids brittle decisions at a single threshold.

Of the $2160$ runs, $1763$ are graded \textsc{Pass}, $280$ \textsc{Marginal} and $117$ \textsc{Fail}.
The rule of Section~\ref{sec:early-termination} stops all $117$ \textsc{Fail} runs, $3$ of the $280$ \textsc{Marginal} runs and no \textsc{Pass} run.
In $100$ further seeds, it again stops all \textsc{Fail} runs and no \textsc{Pass} run.
Against an independent reference, the full-budget grade of another seed of the same circuit and backend, it stops none of the runs whose reference is graded \textsc{Pass} (Appendix~\ref{sec:app-stop}).
Table~\ref{tab:noise_study} lists the six circuits with \textsc{Fail} results.
VQE BeH$_2$, VQE SrH PDM and QML Image Class are detected as unreadable at checkpoint $3$, or at checkpoint $4$ for some seeds of QML Image Class, \ie, after $30\%$ or $40\%$ of their shots.
GHZ-3, GHZ-6 and GHZ-8 stop at median checkpoints of $3$, $4$ and $6$.

\begin{figure*}[t]
    \centering
    \includegraphics[width=0.7\textwidth]{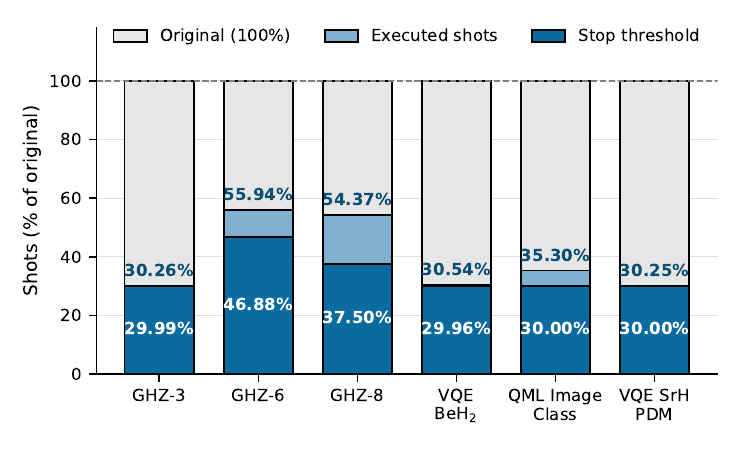}
    \caption{Early-stop effectiveness on the six noise-dominated circuits of Table~\ref{tab:noise_study}, with the stop points preset on the PS. Each bar is normalized to the shot budget of the circuit ($100\%$, gray). The dark blue segment reaches the preset stop point, and the light blue bar gives the shots executed in total (labels, mean of $5$ runs). Its part above the stop point is the shots executed before the stop took effect. The stop points of GHZ-6 and GHZ-8 (checkpoint $5$ of $128$ shots and checkpoint $4$ of $32$ shots) lie at $60$ and $12$ shots, \ie, at $46.88\%$ and $37.50\%$ of their shot budgets. At the stop checkpoints of the simulation instead of the preset ones, the median saving of execution time is $21.77\%$ for GHZ-8 and up to $69.69\%$ for the other circuits (Appendix~\ref{sec:app-stop}).}
    \label{fig:earlystop}
\end{figure*}

On the board, a stop decision reaches the PL with a delay, after which the remaining shots may already have run.
We run the six circuits of Table~\ref{tab:noise_study} on \sys with the stop preset at a checkpoint within the simulated range, $5$ for GHZ-6, $4$ for GHZ-8 and $3$ for the others.
The computation of every checkpoint is emulated with its time measured on the ARM cores of the PS.
This experiment thus measures the response of the stop interface and the time that a stop saves.
The PS starts a checkpoint as soon as the readout data of its last shot are in DRAM.
The uplink delivers these data in half-buffers of $64$ records, \eg, every $4$ shots for GHZ-8, which reads each of its $8$ qubits twice per shot.
All six circuits stop before their shot budget ends (Fig.~\ref{fig:earlystop}).
The stop takes effect $5$ to $24$ shots after the stop point on average and saves $36.97\%$ (GHZ-8) to $69.69\%$ (GHZ-3) of the execution time of the full circuit.
This saving is smaller than the share of shots saved, because the fixed cost of starting the circuit and draining its data remains.
At the stop checkpoints of the simulation, the same delay gives median savings of $21.77\%$ (GHZ-8) to $69.69\%$ (GHZ-3) (Appendix~\ref{sec:app-stop}).
When the PS polls for new readout data continuously instead of every $200~\mu\text{s}$, GHZ-8 stops $2$ shots after the stop point and saves $45.79\%$ of its execution time.

%% file: tables/functionality.tex
\begin{table*}[t]
\centering
\caption{Functionality comparison across three system configurations. \ding{51} indicates that the configuration supports the timing and memory needs of the reconstructed experiment, and 1Q and 2Q denote one- and two-qubit. The \sys entries were executed on the board, and the other entries follow from the command and readout buffer capacities of \qubic.}
\label{tab:capabilities}
\setlength{\aboverulesep}{0pt}
\setlength{\belowrulesep}{0pt}
\renewcommand{\arraystretch}{1.4}
\setlength{\tabcolsep}{4pt}
\scriptsize
\begin{tabular}{l|c|c|c}
\toprule
\rule{0pt}{3ex}{Experiment}
& \makecell{\qubic\\Downlink + Uplink}
& \makecell{\qubic Downlink\\+ \sys Uplink}
& \makecell{\sys\\Downlink + Uplink} \\
\midrule
\makecell[l]{Repeated Ramsey\\($2~\text{ms}$ idle, $5 \!\times\! 10^4$ shots)~\cite{yan2012noise}}                                 & \ding{55} & \ding{51} & \ding{51} \\
\makecell[l]{Charge-Parity Monitoring\\($6~\mu\text{s}$ idle, $8000$ shots)~\cite{riste2013charge}}                               & \ding{55} & \ding{51} & \ding{51} \\
1Q RB ($5101$ Clifford gates)~\cite{tornow2022restless}                                                 & \ding{55} & \ding{55} & \ding{51} \\
2Q RB ($500$ Clifford gates)~\cite{hughes2025trappedion}                                                & \ding{55} & \ding{55} & \ding{51} \\
\bottomrule
\end{tabular}
\end{table*}

%% file: tables/device_endpoints.tex
\begin{table*}[t]
\centering
\caption{Comparison of the \sys and \qubic stacks on the qubit. The table gives the seven pre-registered endpoints~\cite{antq}, their margins, the difference (or ratio) \sys minus (over) \qubic in each pass, and the change of each stack between its two blocks (second minus, or over, first block). The six statistical endpoints carry confidence intervals (CIs), which are $95\%$ except the $90\%$ CIs of the assignment fidelity (two one-sided tests at the $5\%$ level~\cite{schuirmann1987tost}). An endpoint meets the rule when all four intervals lie within the margin. The intervals come from the fit uncertainties (fringe), the binomial error of the held-out shots (assignment fidelity), a bootstrap~\cite{efron1993bootstrap} over the pairs of each block (paired Ramsey) and a bootstrap over the $8$ random sequences, paired across the stacks (EPC of RB~\cite{magesan2011scalable}). The timing loop is an execution check. As pre-registered, both stacks run the same compiled $6.000~\mu\text{s}$ loop and return all $8000$ readouts in every block.}
\label{tab:device_endpoints}
\setlength{\aboverulesep}{0pt}
\setlength{\belowrulesep}{0pt}
\renewcommand{\arraystretch}{1.3}
\setlength{\tabcolsep}{2.5pt}
\scriptsize
\begin{tabular}{l|c|c|c|c|c|c}
\toprule
\rule{0pt}{3ex}Endpoint & Margin & Pass 1 & Pass 2 & \makecell{Change of\\\qubic} & \makecell{Change of\\\sys} & \makecell{Meets\\rule} \\
\midrule
\makecell[l]{Assignment fidelity\\(difference)} & $\pm 0.010$ & \makecell{$+0.0048$\\$[-0.0014, 0.0110]$} & \makecell{$-0.0023$\\$[-0.0085, 0.0038]$} & \makecell{$+0.0064$\\$[+0.0003, 0.0126]$} & \makecell{$-0.0007$\\$[-0.0069, 0.0055]$} & no \\
\makecell[l]{Fringe phase\\(difference, rad)} & $\pm 0.10$ & \makecell{$+0.021$\\$[-0.019, 0.061]$} & \makecell{$+0.022$\\$[-0.020, 0.065]$} & \makecell{$+0.001$\\$[-0.043, 0.045]$} & \makecell{$+0.002$\\$[-0.036, 0.040]$} & yes \\
\makecell[l]{Fringe contrast\\(ratio)} & $0.90$--$1.10$ & \makecell{$1.011$\\$[0.974, 1.049]$} & \makecell{$0.992$\\$[0.953, 1.031]$} & \makecell{$0.996$\\$[0.955, 1.037]$} & \makecell{$0.978$\\$[0.943, 1.012]$} & yes \\
\makecell[l]{Timing loop\\($6~\mu\text{s}$, $8000$ shots)} & identical & identical & identical & identical & identical & yes \\
\makecell[l]{Pair-difference mean\\(difference)} & $\pm 0.02$ & \makecell{$+0.002$\\$[-0.011, 0.014]$} & \makecell{$-0.005$\\$[-0.017, 0.008]$} & \makecell{$+0.002$\\$[-0.010, 0.014]$} & \makecell{$-0.004$\\$[-0.016, 0.008]$} & yes \\
\makecell[l]{Pair-difference\\variance (ratio)} & $0.80$--$1.25$ & \makecell{$1.005$\\$[0.987, 1.023]$} & \makecell{$1.001$\\$[0.984, 1.019]$} & \makecell{$1.013$\\$[0.994, 1.030]$} & \makecell{$1.009$\\$[0.992, 1.028]$} & yes \\
\makecell[l]{EPC of 1Q RB\\(difference)} & $\pm 0.005$ & \makecell{$-0.0003$\\$[-0.0018, 0.0015]$} & \makecell{$-0.0008$\\$[-0.0035, 0.0019]$} & \makecell{$-0.0003$\\$[-0.0014, 0.0013]$} & \makecell{$-0.0007$\\$[-0.0036, 0.0020]$} & yes \\
\bottomrule
\end{tabular}
\end{table*}

%% file: tables/physics.tex
\begin{table*}[t]
\centering
\caption{What the four benchmark experiments gain from \sys, on the bench and on the qubit, with the measured numbers (Sections~\ref{sec:eval-rq1} to~\ref{sec:eval-noise}). ``Model'' marks a number that also depends on the stated noise model, and ``simulation'' marks a number from simulated records.}
\label{tab:physics}
\setlength{\aboverulesep}{0pt}
\setlength{\belowrulesep}{0pt}
\renewcommand{\arraystretch}{1.2}
\scriptsize
\begin{tabularx}{\textwidth}{>{\raggedright\arraybackslash}p{0.19\textwidth}|>{\raggedright\arraybackslash}p{0.055\textwidth}|>{\raggedright\arraybackslash}X|>{\raggedright\arraybackslash}X}
\toprule
\rule{0pt}{2.5ex}Experiment & Where & \qubic baseline & \sys \\
\midrule
\multirow{4}{=}{(1) Repeated Ramsey for $1/f$ noise ($2~\text{ms}$ idle)~\cite{yan2012noise}} & Bench & Not supported & Every shot at the programmed period ($10$ records of about $500$ shots) \\
\cline{2-4}
 & Qubit & Records of $25000$ pairs in $2.048~\text{s}$ runs with untimed gaps, spectrum per run from $0.49~\text{Hz}$ & One uninterrupted $100~\text{s}$ record with a frequency spacing of $0.01~\text{Hz}$ \\
 & & Upper limit on $\sqrt{S_f(1~\text{Hz})}$, run by run, of $22.3$ and $21.8~\text{kHz}/\sqrt{\text{Hz}}$ for its records, $22.5$ and $25.3~\text{kHz}/\sqrt{\text{Hz}}$ for the \sys records (model) & $12.6$ and $17.7~\text{kHz}/\sqrt{\text{Hz}}$ (model) \\
 & & $50\%$ detection level of the same records in $2.048~\text{s}$ runs, $12.3$ and $16.7~\text{kHz}/\sqrt{\text{Hz}}$ (simulation) & $10.5$ and $13.1~\text{kHz}/\sqrt{\text{Hz}}$, $1.37$ and $1.62$ times lower spectral density (simulation) \\
\midrule
\multirow{4}{=}{(2) Charge-parity detection ($6~\mu\text{s}$ idle, $8000$ samples)~\cite{riste2013charge}} & Bench & Not supported & Every shot at the programmed period ($3$ records of $8000$ shots) \\
\cline{2-4}
 & Qubit & Timing loop in runs of at most $1024$ shots, with host round trips between them that break the fixed spacing & One program with all $8000$ shots at the compiled spacing of $6.000~\mu\text{s}$ \\
\midrule
\multirow{2}{=}{(3) 1Q RB, $5101$ Cliffords~\cite{tornow2022restless}} & Bench & Not supported & $5$ sub-circuits per shot, no boundary wait above $16~\text{ns}$, run time $0.34\%$ above the programmed one \\
\cline{2-4}
 & Qubit & Up to about $1000$ Cliffords ($1024$ exceed the $2048$ commands) & $1024$, $2048$ and $5101$ Cliffords in $24$ runs. Where both run, EPC difference $+0.0003$ ($[-0.0001, 0.0007]$). Beyond $1024$, the decay has saturated \\
\midrule
(4) 2Q RB, $500$ Cliffords~\cite{hughes2025trappedion} & Bench & Not supported & $3$ sub-circuits per shot, no boundary wait above $16~\text{ns}$, run time $0.11\%$ above the programmed one \\
\midrule
A handover inside a circuit & Qubit & --- & Acts as an idle of $144~\text{ns}$, with phase $+0.038~\text{rad}$ ($[-0.004, 0.082]$) and contrast ratio $1.002$ ($[0.967, 1.040]$) against the idle control. At most $3 \times 10^{-5}$ per Clifford in long RB (model) \\
\bottomrule
\end{tabularx}
\end{table*}

%% file: tables/speedup.tex
\begin{table*}[t]
\centering
\caption{Excess run time over the baseline execution for one single circuit. ``Q \#'' and ``Shot \#'' denote the numbers of qubits and shots. The per-shot time is the duration of one shot measured on the \qubic baseline bitfile. The reference time (Ref.) is then calculated as $\text{Shot \#} \times \text{Per-Shot time}$. No quantum device is connected on the bench. For each configuration, the columns give the measured execution time (ms) and the excess run time $\Delta$ over the reference time, in ms and as a percentage of the reference time.}
\label{tab:speedup}
\setlength{\aboverulesep}{0pt}
\setlength{\belowrulesep}{0pt}
\renewcommand{\arraystretch}{1.3}
\setlength{\tabcolsep}{2.2pt}
\scriptsize
\begin{tabular}{l|c|c|c|c|rrr|rrr|rrr}
\toprule
\rule{0pt}{3ex}\multirow{2}{*}[-1ex]{Name} & \multirow{2}{*}[-1ex]{Q \#} & \multirow{2}{*}[-1ex]{Shot \#} & \multirow{2}{*}[-1ex]{\makecell[c]{Per-Shot\\$(\mu\text{s})$}} & \multirow{2}{*}[-1ex]{\makecell[c]{Ref.\\$(\text{ms})$}}
& \multicolumn{3}{c|}{\makecell{\qubic\\Downlink + Uplink}}
& \multicolumn{3}{c|}{\makecell{\qubic Downlink\\+ \sys Uplink}}
& \multicolumn{3}{c}{\makecell{\sys\\Downlink + Uplink}} \\
\cline{6-8}\cline{9-11}\cline{12-14}
 & & & & & $(\text{ms})$ & $\Delta$ $(\text{ms})$ & $\Delta$ $(\%)$ & $(\text{ms})$ & $\Delta$ $(\text{ms})$ & $\Delta$ $(\%)$ & $(\text{ms})$ & $\Delta$ $(\text{ms})$ & $\Delta$ $(\%)$ \\
\midrule
QAOA MaxCut~\cite{hao2024qaoa} & $12$ & $10000$ & $43.39$ & $433.87$ & $592.69$ & $158.83$ & $36.61\%$ & $437.08$ & $3.22$ & $0.74\%$ & $434.10$ & $0.23$ & $0.05\%$ \\
Bell state (GHZ-2)~\cite{baumer2021scalable} & $2$ & $24576$ & $8.75$ & $215.05$ & $402.46$ & $187.41$ & $87.15\%$ & $217.43$ & $2.38$ & $1.11\%$ & $214.83$ & $-0.21$ & $-0.10\%$ \\
VQE BeH2~\cite{belaloui2025vqe} & $14$ & $4096$ & $46.57$ & $190.73$ & $266.37$ & $75.64$ & $39.66\%$ & $193.68$ & $2.94$ & $1.54\%$ & $190.92$ & $0.18$ & $0.10\%$ \\
VQE SrH PDM~\cite{chawla2024srh} & $12$ & $4000$ & $43.01$ & $172.02$ & $236.34$ & $64.31$ & $37.39\%$ & $174.81$ & $2.79$ & $1.62\%$ & $172.17$ & $0.15$ & $0.09\%$ \\
GHZ-4~\cite{baumer2021scalable} & $4$ & $8192$ & $15.00$ & $122.87$ & $200.57$ & $77.70$ & $63.24\%$ & $125.34$ & $2.47$ & $2.01\%$ & $122.94$ & $0.07$ & $0.05\%$ \\
GHZ-3~\cite{baumer2021scalable} & $3$ & $8192$ & $11.86$ & $97.19$ & $158.93$ & $61.74$ & $63.52\%$ & $99.64$ & $2.45$ & $2.52\%$ & $97.29$ & $0.10$ & $0.10\%$ \\
VQE 1Q Eigensolver~\cite{grove_vqe} & $1$ & $10000$ & $5.59$ & $55.93$ & $114.40$ & $58.47$ & $104.54\%$ & $58.16$ & $2.23$ & $3.99\%$ & $55.93$ & $0.00$ & $0.00\%$ \\
Grover~\cite{vemula2022grover} & $6$ & $1024$ & $59.64$ & $61.07$ & $72.62$ & $11.55$ & $18.91\%$ & $64.43$ & $3.36$ & $5.51\%$ & $61.14$ & $0.07$ & $0.12\%$ \\
VQE HeH+~\cite{liang2024aishot} & $2$ & $4000$ & $8.75$ & $35.00$ & $66.01$ & $31.01$ & $88.59\%$ & $37.28$ & $2.28$ & $6.50\%$ & $35.00$ & $0.00$ & $0.00\%$ \\
Bernstein--Vazirani~\cite{ibm_query_algorithms} & $5$ & $1024$ & $19.09$ & $19.55$ & $29.96$ & $10.42$ & $53.29\%$ & $21.94$ & $2.39$ & $12.22\%$ & $19.61$ & $0.06$ & $0.30\%$ \\
GHZ-5~\cite{baumer2021scalable} & $5$ & $1024$ & $18.13$ & $18.57$ & $28.28$ & $9.71$ & $52.31\%$ & $20.94$ & $2.37$ & $12.74\%$ & $18.63$ & $0.06$ & $0.34\%$ \\
Shor $N{=}15$~\cite{amico2019shor} & $5$ & $1000$ & $17.82$ & $17.82$ & $26.75$ & $8.93$ & $50.13\%$ & $20.18$ & $2.36$ & $13.26\%$ & $17.89$ & $0.07$ & $0.38\%$ \\
QSVM Radar~\cite{agnihotri2026qsvm} & $4$ & $1024$ & $17.29$ & $17.70$ & $27.35$ & $9.65$ & $54.50\%$ & $20.09$ & $2.39$ & $13.48\%$ & $17.75$ & $0.05$ & $0.28\%$ \\
QML Image Class~\cite{gharibyan2025qml} & $14$ & $400$ & $50.29$ & $20.12$ & $28.01$ & $7.90$ & $39.27\%$ & $22.85$ & $2.73$ & $13.58\%$ & $20.27$ & $0.16$ & $0.78\%$ \\
Quantum Teleportation~\cite{ibm_teleportation} & $3$ & $1000$ & $14.65$ & $14.65$ & $23.07$ & $8.42$ & $57.47\%$ & $16.98$ & $2.33$ & $15.91\%$ & $14.70$ & $0.05$ & $0.35\%$ \\
Deutsch--Jozsa~\cite{ibm_query_algorithms} & $3$ & $1024$ & $12.20$ & $12.50$ & $19.49$ & $6.99$ & $55.95\%$ & $14.85$ & $2.35$ & $18.82\%$ & $12.56$ & $0.07$ & $0.53\%$ \\
QFT~\cite{teegarden2026threemonths} & $4$ & $500$ & $26.63$ & $13.32$ & $18.15$ & $4.83$ & $36.28\%$ & $15.89$ & $2.57$ & $19.30\%$ & $13.37$ & $0.06$ & $0.43\%$ \\
QAOA $p{=}1$~\cite{oqd_qaoa} & $2$ & $1000$ & $9.11$ & $9.11$ & $16.80$ & $7.69$ & $84.45\%$ & $11.40$ & $2.29$ & $25.17\%$ & $9.13$ & $0.02$ & $0.27\%$ \\
GHZ-6~\cite{baumer2021scalable} & $6$ & $128$ & $21.25$ & $2.72$ & $6.19$ & $3.47$ & $127.62\%$ & $5.10$ & $2.38$ & $87.42\%$ & $2.80$ & $0.08$ & $2.90\%$ \\
GHZ-8~\cite{baumer2021scalable} & $8$ & $32$ & $27.52$ & $0.88$ & $3.64$ & $2.76$ & $313.04\%$ & $3.30$ & $2.42$ & $274.41\%$ & $0.98$ & $0.10$ & $11.08\%$ \\
\bottomrule
\end{tabular}
\end{table*}

%% file: tables/speedup_batch.tex
\begin{table*}[t]
\centering
\caption{Batch benchmark results for $30$-circuit batches. For each setting we report the total time and the excess run time $\Delta$ over the reference time (Ref.), in ms and as a percentage of the reference time. In the last row, the two \qubic Downlink configurations are compiled against the shared table layout that \sys uses in every row. The reference time of a batch is the sum of those of its circuits (Table~\ref{tab:speedup}). The random row averages $10$ random batches of $30$ circuits.}
\label{tab:batch_speedup}
\setlength{\aboverulesep}{0pt}
\setlength{\belowrulesep}{0pt}
\renewcommand{\arraystretch}{1.4}
\setlength{\tabcolsep}{2.3pt}
\scriptsize
\begin{tabular}{l|c|rrr|rrr|rrr}
\toprule
\rule{0pt}{3ex}\multirow{2}{*}[-1ex]{Name} & \multirow{2}{*}[-1ex]{\makecell[c]{Ref.\\$(\text{ms})$}}
& \multicolumn{3}{c|}{\makecell{\qubic\\Downlink + Uplink}}
& \multicolumn{3}{c|}{\makecell{\qubic Downlink\\+ \sys Uplink}}
& \multicolumn{3}{c}{\makecell{\sys\\Downlink + Uplink}} \\
\cline{3-5}\cline{6-8}\cline{9-11}
 & & $(\text{ms})$ & $\Delta$ $(\text{ms})$ & $\Delta$ $(\%)$ & $(\text{ms})$ & $\Delta$ $(\text{ms})$ & $\Delta$ $(\%)$ & $(\text{ms})$ & $\Delta$ $(\text{ms})$ & $\Delta$ $(\%)$ \\
\midrule
GHZ-8 $\times$ 30 & $26.42$ & $108.97$ & $82.55$ & $312.49\%$ & $105.60$ & $79.18$ & $299.70\%$ & $26.65$ & $0.23$ & $0.87\%$ \\
QAOA MaxCut $\times$ 30 & $13016.01$ & $17788.56$ & $4772.55$ & $36.67\%$ & $13198.87$ & $182.86$ & $1.40\%$ & $13019.11$ & $3.10$ & $0.02\%$ \\
Random $30$-circuit avg & $2520.76$ & $4009.81$ & $1489.05$ & $59.07\%$ & $2849.94$ & $329.18$ & $13.06\%$ & $2520.68$ & $-0.07$ & $0.00\%$ \\
\makecell[l]{Random $30$-circuit avg,\\shared tables} & $2520.76$ & $3794.18$ & $1273.42$ & $50.52\%$ & $2620.92$ & $100.16$ & $3.97\%$ & $2520.68$ & $-0.07$ & $0.00\%$ \\
\bottomrule
\end{tabular}
\end{table*}

%% file: tables/noise_analysis.tex
\newcolumntype{C}{>{\centering\arraybackslash}X}

\begin{table*}[t]
\centering
\caption{Readability of benchmark circuits on IBM noisy simulator backends of Qiskit~\cite{Qiskit} over $20$ simulator seeds, for the circuits with FAIL results. A run is graded by the Hellinger fidelity (HF)~\cite{hartnett2024learning} and the total variation distance (TVD)~\cite{dahlhauser2021modeling} of its output distribution as \textsc{Pass} if $\mathrm{HF} > 0.7$ and $\mathrm{TVD} < 0.3$, \textsc{Fail} if $\mathrm{HF} \le 0.4$ and $\mathrm{TVD} \ge 0.6$, and \textsc{Marginal} otherwise. A cell gives the grade of all $20$ seeds, or the number of seeds of each grade when they differ. ``Stop at'' gives the median (range) of the checkpoint at which the early-termination rule stops the FAIL runs. ``—'' indicates the circuit was not able to run on that backend.}
\label{tab:noise_study}
\setlength{\aboverulesep}{0pt}
\setlength{\belowrulesep}{0pt}
\renewcommand{\arraystretch}{1.8}
\setlength{\tabcolsep}{2pt}
\scriptsize
\begin{tabularx}{\textwidth}{l|c|>{\hsize=0.75\hsize}C|>{\hsize=1.2\hsize}C|>{\hsize=1.0\hsize}C|>{\hsize=0.75\hsize}C|>{\hsize=0.9\hsize}C|>{\hsize=1.4\hsize}C}
\toprule
\rule{0pt}{3ex}Circuit & Stop at & \textit{Manila} & \textit{Lagos} & \textit{Guadalupe} & \textit{Algiers} & \textit{Sherbrooke} & \textit{Torino} \\
\midrule
GHZ-3                 & 3 (3--9)  & PASS & 17 FAIL,\newline 3 MARGINAL & PASS & PASS & PASS & PASS \\
GHZ-6                 & 4 (3--7)  & —    & FAIL & PASS & PASS & PASS & 10 PASS,\newline 10 MARGINAL \\
GHZ-8                 & 6 (3--10) & —    & —    & PASS & PASS & FAIL & MARGINAL \\
VQE BeH$_2$           & 3         & —    & —    & MARGINAL & PASS & FAIL & MARGINAL \\
QML Image Class       & 3 (3--4)  & —    & —    & MARGINAL & PASS & FAIL & MARGINAL \\
VQE SrH PDM           & 3         & —    & —    & MARGINAL & PASS & FAIL & MARGINAL \\
\bottomrule
\end{tabularx}
\end{table*}

%% file: sections/discuss.tex
\section{Limitations and Future Work}
\label{sec:future_work}
The current \sys has limits in its host link, its early-termination monitor, its support for branches, its scale and its validation on qubits.
We discuss each limit below together with our plan to remove it.

\textbf{Host link.}
In the current implementation, the PS serves as a proxy between the host machine and the PL.
As the DRAM and BRAM bandwidth scale up, the TCP/IP link between the PS and the host, capped at about $120~\text{MB/s}$ on the ZCU216, becomes the new bottleneck.
This link sets the host term of the refill condition (Section~\ref{sec:eval-rq1}).
Under this term, each circuit of a batch much larger than what DRAM and the PS hold must run about $0.27~\text{ms}$ per DSP core it uses.
The link also limits how quickly early termination responds.
All batches in our evaluation fit into DRAM and the PS circuit buffer, so we have not measured this term.
We plan to bypass the PS and connect the PL directly to the host through a $100$~Gbps Ethernet interface.
The PS servers now move the commands into DRAM, write the metadata of each circuit into the PL and read the ring buffer, and these tasks will then move to the host and the PL.

\textbf{Live early termination.}
The early-termination monitor does not yet run live.
In our evaluation, the stop checkpoint of each circuit is chosen within the range given by the noise simulation study and preset in the stream server on the PS, and the cost of each checkpoint is emulated with its time measured on the PS (Section~\ref{sec:eval-stop}).
We plan to run the monitor as a live service on the PS, which evaluates each checkpoint as the readout data arrive and sends the \Code{Stop} signal through the same path.

\textbf{Branches and feed-forward.}
The current \sys design assumes sequential command consumption within each ping-pong half, and the compiler splits only straight-line programs, with a barrier across all cores at every cut (Section~\ref{sec:downlink}).
This design limits mid-circuit measurement and feed-forward control.
When a measurement outcome selects a branch whose commands lie outside the active half, the DSP must stall while the target is fetched from DRAM or, in the worst case, from the host.
The cost of this stall grows with the memory level that holds the target.
We plan to remove this stall with a predictive preload mechanism.
The mechanism annotates each conditional branch with a likely target and speculatively stages it one level closer to the DSP, from the host to DRAM or from DRAM to the inactive BRAM half, before the measurement resolves.
The prediction can follow the empirical branch distributions of earlier shots, and on-demand fetching serves as a fallback when a prediction misses.

\textbf{Multiple boards.}
Our design is limited to a single ZCU216 board, which drives at most $14$ qubits.
Scaling to wider circuits requires board-to-board clock synchronization and low-latency cross-board communication.
We plan to integrate the solution of~\cite{xu2025multifpga} with \sys and design a distributed memory hierarchy, in which the BRAM and DRAM tiers of all boards act as a single logical pool.
In this design, the drive and readout channels of a qubit can be assigned to different boards, which enables wider circuits and balances the load across boards when the readout traffic is uneven.

\textbf{Validation on qubits.}
Our evaluation on a superconducting qubit covers one qubit of one device, measured in one session with the $8$-core bitfile.
The comparison of the stacks also departs from its pre-registered plan in the block order and in the omitted transport controls (Appendix~\ref{sec:app-device}), which limits how far the measured differences can be attributed to the stacks.
The evaluation does not yet show a resolved noise spectrum, which on this qubit is limited by the binomial noise of the single-shot outcomes, or a gain in fidelity estimates.
We have not yet measured how longer uninterrupted records improve $1/f$ noise spectroscopy on a qubit with a resolvable $1/f$ component.
Nor have we measured how long RB sequences sharpen fidelity estimates on devices whose error per Clifford is low enough for the decay to stay resolvable at such depths.
We have also not tested the complete measurement chain with a frequency modulation of known spectrum injected through the drive.
We plan to extend the validation to several qubits, to streamed two-qubit RB on a device and to charge-parity monitoring on a qubit that is sensitive to offset charge.
We also plan to obtain these end-to-end precision numbers on calibrated transmon devices at the Advanced Quantum Testbed (AQT)~\cite{aqt} at Lawrence Berkeley National Laboratory.

%% file: sections/related.tex
\section{Related Work}
\label{sec:related}
The works most relevant to ours place DRAM in quantum control systems, optimize other layers of the control stack, scale control systems across boards, or reduce the number of measurement shots.

\subsection{DRAM in Quantum Control Systems}
Several control systems already attach DRAM to the FPGA.
ARTIQ~\cite{artiq_dma} records a sequence of real-time events into DRAM once and replays it by DMA, which removes the CPU from the event path of that sequence.
QICK~\cite{qick_ddr4} provides a DDR4 buffer that captures long streams of raw ADC samples for offline analysis.
Dudley et al.~\cite{dudley2025sgdma} store the gate sequences of an RFSoC control system for trapped ions in DDR4 memory and move them to a state machine in the programmable logic by scatter-gather DMA.
On the bench, they measure transfer latencies of $849$ to $1516~\text{ns}$ and the throughput for different descriptor and payload sizes.
In these systems, DRAM holds one kind of data, either a stored control sequence or raw samples.
QICK also streams accumulated readout during a run.
A thread on the PS reads its readout buffers of $16384$ entries while the program executes~\cite{qick_streamer, qick_ddr4}, a role that the \sys Uplink plays on \qubic with DRAM behind the buffers.
\sys places DRAM in both the command and the readout path of every circuit and hides its non-deterministic latency with ping-pong BRAM buffers.
As a result, circuits whose commands exceed the on-chip memory run with a constant handover, and circuits are loaded, executed and read out as a pipeline.
Table~\ref{tab:compare} compares these capabilities as far as the cited sources describe them.

\input{tables/compare}

\subsection{Performance Optimizations Across the Quantum Control Stack}
Other works optimize different layers of the control stack.
Qtenon~\cite{qtenon} co-integrates a RISC-V host and a quantum controller on the same ASIC through a unified memory hierarchy and a custom instruction set on the Rocket custom coprocessor (RoCC) interface.
By cutting the communication latency between host and controller, it reports up to $14.9\times$ speedup on variational workloads.
Its focus is this latency rather than the capacity of the program memory of the controller.
UQP~\cite{uqp} lowers programs in the quantum intermediate representation (QIR) to a hybrid instruction set and extends the HiSEP-Q controller with modality-specific blocks.
One quantum control processor can thus drive multiple qubit technologies without recompilation overhead.
\sys runs on commodity RFSoCs and attacks the on-chip program-capacity bottleneck directly.
The three works improve the efficiency of quantum execution along orthogonal axes of the control stack and are complementary.

\subsection{Multi-FPGA Scaling for Quantum Control}
Two recent efforts extend RFSoC-based control systems across multiple boards to scale the qubit count.
Xu et al.~\cite{xu2025multifpga} extend QubiC with Precision Time Protocol clock synchronization and an Aurora 64B/66B fiber link for low-latency readout and feed-forward transfer across ZCU216 boards.
Manarat~\cite{silva2026manarat} takes a parallel approach in the QICK ecosystem.
It uses a shared HMC7044 reference and a wired-AND synchronization bus to align the execution of the tProcessor to below $100$~ps across boards, and it is validated on a $10$-qubit flux-tunable processor.
These works and \sys address orthogonal dimensions of circuit scaling and can be combined.
They increase circuit \emph{width} by coordinating more qubits across boards, whereas \sys increases circuit \emph{depth} on each board by breaking the BRAM capacity bottleneck that caps commands and shots.

\subsection{Reducing Measurement Shots}
\sys's early-termination mechanism complements prior work on reducing measurement shots, which works at two layers.
At the host layer, iCANS~\cite{kubler2020adaptive} adaptively allocates shots across gradient components in variational optimization.
It spends fewer shots on directions far from the optimum and more on those near convergence.
QuFid~\cite{adaptive_fidelity_qufid} models a circuit as a directed acyclic graph and uses a noise-propagation operator with a confidence-interval stopping rule to terminate measurement once the fidelity has converged.
At the control-system layer, \sys reduces shots from a different angle.
It streams readout as shots complete, so a circuit overwhelmed by noise can be terminated during execution instead of after the full batch returns to the host.
The two layers are complementary.
Host-layer policies decide \emph{how many} shots to request, and \sys decides \emph{when within a batch} to stop delivering them, so combining them reduces shots along both axes.

%% file: tables/compare.tex
\begin{table*}[t]
\centering
\caption{Capabilities of \sys and of related control systems as described in the cited sources. \ding{51} marks a described capability, --- one not described in the cited source, and \ding{55} one not supported.}
\label{tab:compare}
\setlength{\aboverulesep}{0pt}
\setlength{\belowrulesep}{0pt}
\renewcommand{\arraystretch}{1.3}
\footnotesize
\begin{tabularx}{\textwidth}{>{\raggedright\arraybackslash}X|c|c|c|c|c}
\toprule
\rule{0pt}{2.5ex}Capability & \makecell{\qubic\\baseline~\cite{xu2023qubic}} & \makecell{ARTIQ\\DMA~\cite{artiq_dma}} & \makecell{QICK~\cite{qick_ddr4,qick_streamer}} & \makecell{RFSoC\\SG-DMA~\cite{dudley2025sgdma}} & \sys \\
\midrule
Control sequences held in DRAM and fed to the real-time logic during a run & \ding{55} & \ding{51} & --- & \ding{51} & \ding{51} \\
Readout results streamed to the host while the program runs & \ding{55} & --- & \ding{51} & --- & \ding{51} \\
Switch between stored segments on all cores at one boundary & \ding{55} & --- & --- & --- & \ding{51} \\
Measured cost of that switch & \ding{55} & --- & --- & --- & \ding{51} \\
Phase state carried across segments and checked on a qubit & \ding{55} & --- & --- & --- & \ding{51} \\
Batches of different circuits with shared tables & \ding{55} & --- & --- & --- & \ding{51} \\
Interface that stops a running circuit & \ding{55} & --- & --- & --- & \ding{51} \\
\bottomrule
\end{tabularx}
\end{table*}

%% file: sections/conclude.tex
\section{Conclusion}

Quantum control boards keep circuits and readout results in on-chip BRAM, whose small capacity caps the depth and length of experiments and forces a punched-card workflow.
DRAM offers the capacity, but its latency is not deterministic.
In this work, we presented \sys, a memory-hierarchy design that breaks the BRAM capacity bottleneck of quantum control boards by using DRAM as the main memory pool and repurposing BRAM as an explicitly managed staging buffer.
With ping-pong buffering and decoupled downlink and uplink data paths, \sys turns the punched-card workflow into a pipeline and keeps the inter-circuit timing deterministic whenever each circuit outlasts the refill of the next.
Each switch to a ready command buffer takes a constant $134~\text{ns}$, only $14~\text{ns}$ more than the restart of the DSP core between shots.
On the ZCU216, \sys supports deep randomized benchmarking and long-duration noise experiments that are infeasible under \qubic.
For computational circuits with at least $1000$ shots, it reduces the time the board spends beyond the circuit execution to at most $0.53\%$ of the execution time, and it improves the board-level throughput of $30$-circuit batches by $51.06\%$ on average against a baseline with the same shared tables ($59.90\%$ against the stock baseline).
On a superconducting qubit, \sys executes a $5101$-Clifford randomized benchmarking sequence, gives Ramsey and randomized benchmarking results equivalent to the \qubic baseline within pre-registered margins, and acquires uninterrupted $100~\text{s}$ Ramsey records whose noise spectra show no resolved $1/f$ component.
These results show that DRAM, carefully integrated, can enter the real-time control path of FPGA-based quantum control systems and enable both deeper experiments and higher-throughput computing on commodity RFSoC hardware.

%% file: sections/appendix.tex
\section{Timing of the command path}
\label{sec:app-timing}
The two handshake flags of a ping-pong buffer, \Code{Write Done} and \Code{Read Done}, cross between the $333~\text{MHz}$ DRAM side and the $500~\text{MHz}$ DSP side through toggle-based pulse synchronizers, and the fill state of each half crosses through two-stage synchronizers.
Table~\ref{tab:handover} lists the events of one handover in our cycle-accurate simulation of the command path around the \qubic DSP core~\cite{antq}.
The simulation uses the released RTL of the ping-pong buffers, the command writer, the flag counter, the synchronizers, the DSP cores and the processor.
Its board top and DSP wrapper differ from the release only in removed monitor paths of the ADC and DAC.
Its command sequencer differs only in how a batch starts and ends, with a level-sensitive start and automatic re-arming instead of a start pulse, and in two status outputs.
The transitions between circuits inside a batch are therefore those of the release.
The handover takes $67$ cycles ($134~\text{ns}$) from the last sample of one circuit to the first sample of the next, less the programmed start time.
Between two shots of one program, which pass through neither the sequencer nor the buffer switch, the same interval takes $60$ cycles ($120~\text{ns}$).

\input{tables/handover}

The gateware counts the DSP cycles between the release of the current halves and the moment the next halves of all cores are ready.
When the next halves are ready before the boundary, the count is $1$ cycle and the handover follows the fixed sequence of Table~\ref{tab:handover}.
A half that becomes ready only during the boundary lengthens the handover cycle for cycle.
A sticky flag is set when the count exceeds $8$ cycles ($16~\text{ns}$), a margin for synchronizing the fill state across the clock domains, and the counter keeps the value of the last boundary of a run.
In all $429$ streamed runs of this paper, on the bench and on the qubit, the flag is clear and the counter reads $1$ cycle (Fig.~\ref{fig:timing}b).
The last handover of each run thus followed the simulated sequence to the cycle, and every other boundary took at most $7$ extra cycles ($14~\text{ns}$).
These runs include the batches of Section~\ref{sec:eval-rq2}, which stream commands and readout at the same time on $1$ to $14$ cores.
Above the threshold, the runs of the refill sweep also end with the nominal wait of one cycle.
A boundary can wait between $1$ and $7$ extra cycles only if the fill of its next halves completes within $16~\text{ns}$ of the boundary.
In (3) and (4), the fills complete $43.5$ and $72.2~\mu\text{s}$ before the boundary at the fitted rate.
The counter covers the only part of a handover that depends on the command supply, and the oscilloscope measures the whole handover on one program.
Over $12$ captures each, the median gap between the two marker pulses is $545~\text{ns}$ (standard deviation $0.6~\text{ns}$) with a handover between them and $401~\text{ns}$ ($0.7~\text{ns}$) without, so the captures resolve no additional spread from the handover.

Every sub-circuit start except the first of a run counts as a boundary.
Experiment (3) runs $5$ sub-circuits per shot for $1024$ shots, and (4) runs $3$ per shot for $5000$ shots.
These runs thus have $5119$ and $14999$ boundaries, of which $4$ and $2$ lie inside each shot and one lies between consecutive shots.
On the oscilloscope, the shot period of (3) is $0.71~\mu\text{s}$ ($0.28\%$) longer than programmed, which is the cost of its five handovers per shot ($5 \times 144~\text{ns} = 0.72~\mu\text{s}$).
The rest of its run-time excess of $0.34\%$ is the fixed cost of starting the run and draining its data.

A cut inside a circuit acts as a barrier across all cores.
For example, suppose that core $0$ plays a $200~\text{ns}$ pulse that begins $100~\text{ns}$ before the cut, while core $1$ begins a pulse at the cut.
The split program then starts the pulse of core $1$ only after the pulse of core $0$ has ended, $244~\text{ns}$ instead of $144~\text{ns}$ later than the unbroken program.
The compiler warns about such a cut and reports the overlap of $100~\text{ns}$.
A cut at an idle point keeps the relative timing of all cores and delays everything after it by $144~\text{ns}$.
Every cut in our data lies at an idle point.
The compiler split (4), and the tool of an earlier release (v1.8) of~\cite{antq} split (3) and the device programs, which drive a single qubit, between gates, so no pulse plays across any of their cuts.

The refill condition of Section~\ref{sec:eval-rq1} is written for a batch of equal circuits.
In a batch of different circuits, the first term applies to every transition.
Each circuit must then last at least $t_0 + n' S/B_D$, where $n'$ is the number of cores of the circuit that follows it, and the averaged second and third terms do not ensure this for each circuit.
The value $N_P = 146$ corresponds to the default circuit buffer of $64~\text{MiB}$ on the PS.
The software does not check the condition before a run.
A transition that misses it still executes every command in order, but the DSP waits at the boundary until the commands arrive, which lengthens that handover by the wait.
The run is not rejected, and the flag and the wait at the last boundary are returned with its results (Table~\ref{tab:contract}).

We fit $B_D$ and $t_0$ to the thresholds measured by sweeping the circuit duration on both \sys bitfiles.
On the $14$-core bitfile, these thresholds lie between $52.7$ and $53.4~\mu\text{s}$ for $n = 14$ and between $5.3$ and $5.9~\mu\text{s}$ for $n = 1$.
The fit reproduces all four measured thresholds, for all cores and for one core on each bitfile, within the $0.6~\mu\text{s}$ step of the sweep.
Above its threshold, every measured batch runs without a stall.
Below it, the wait at the last boundary shrinks one for one as the circuit grows, for $n = 14$ from $51.6~\mu\text{s}$ without added duration to $1.7~\mu\text{s}$ with $50~\mu\text{s}$ added (Fig.~\ref{fig:timing}c).
The fitted thresholds describe the tested configurations and do not bound the worst-case service time of the DRAM.
For a given run, the flag gives the bound, since a clear flag shows that no boundary of the run waited more than $16~\text{ns}$.
We measure $B_P = 5.19~\text{GB/s}$ from the upload times of the circuits in Table~\ref{tab:speedup}, and $B_H \approx 0.12~\text{GB/s}$ is the rate of the TCP link between the PS and the host (Section~\ref{sec:future_work}).
The shortest sub-circuits, those of (3) and (4), last $8.7$ and $8.8$ times their thresholds.

\section{Comparison of the two stacks on the qubit}
\label{sec:app-device}
The measured qubit had $T_1 = 48~\mu\text{s}$ and $T_2^* = 1.4~\mu\text{s}$ in the calibration of the session day.
A survey of the device on the same day gave $T_2^*$ between $0.7$ and $1.6~\mu\text{s}$ on the six qubits where it could be fitted.
The four blocks (\qubic, \sys, \qubic, \sys) form two passes of one block per stack.
For each of the six statistical endpoints of Table~\ref{tab:device_endpoints} and each pass, we compute the difference between \sys and \qubic (\sys minus \qubic, or their ratio for the contrast and the variance) with its CI.
The CI is $95\%$ for all endpoints except the assignment fidelity.
The pre-registered test of the assignment fidelity is a pair of one-sided tests at the $5\%$ level~\cite{schuirmann1987tost}, so it uses the $90\%$ CI.
The pre-registered rule regards an endpoint as equivalent only when the CIs of both passes and of the change of each stack between its own two blocks lie within the margin.
The timing loop is an execution check.
Both stacks run the same compiled loop of $6.000~\mu\text{s}$ and return all $8000$ readouts in every block.
\sys runs the loop as one program with fixed shot spacing, and \qubic, whose readout buffer holds $1024$ records, runs it in runs of at most $1024$ shots with host round trips in between.

The assignment fidelity does not meet the rule.
Its second pass lies inside the margin of $\pm 0.010$ ($-0.0023$, 90\% CI $[-0.0085, 0.0038]$).
However, the first pass ($+0.0048$, $[-0.0014, 0.0110]$) and the change of the \qubic baseline between its two blocks ($+0.0064$, $[+0.0003, 0.0126]$) reach beyond the margin at the upper end.
The baseline thus changed between its own blocks by as much as the two stacks differ.
The EPC margin of $\pm 0.005$ was set for the EPC of $0.01$ to $0.03$ that we expected before the session.
The measured EPC is $0.0034$ (\qubic) and $0.0031$ (\sys) in the first pass and $0.0032$ and $0.0024$ in the second, so the margin is wider than the EPC itself.
The CIs of the difference, $[-0.0018, 0.0015]$ and $[-0.0035, 0.0019]$, bound it more tightly than the margin, but the second-pass interval still allows an increase of $0.0019$, more than half of the EPC.
The comparison over the RB depths of Fig.~\ref{fig:device}a bounds the difference to $[-0.0001, 0.0007]$ (Appendix~\ref{sec:app-handover}).

The pre-registration~\cite{antq} states the endpoints, margins and decision rule, and the reported analysis departs from it in four points.
First, the blocks ran in the order \qubic, \sys, \qubic, \sys instead of the planned A-B-B-A order.
Second, each block has one classifier, trained on the first half and evaluated on the second half of its preparation shots.
The cross-applied classifiers, the stack-against-itself transport control, the $T_1$ sweep and the descriptive statistics of the readout clouds in the plan are not part of the reported blocks.
Third, the margin of the assignment fidelity is applied to the difference of each pass with its $90\%$ CI, instead of to averages over paired blocks.
Fourth, the RB of each block uses $5$ lengths ($2$ to $32$ Cliffords), $8$ random sequences and $256$ shots per sequence, and the floor of its fit comes from the readout calibration of the block.

\section{Handover inside a circuit and deep RB}
\label{sec:app-handover}
Against the unbroken program, the handover multiplies the fringe contrast by $0.903$ (95\% CI $[0.874, 0.935]$).
This factor matches the $0.902$ that an exponential $T_2^*$ decay predicts for the $144~\text{ns}$ that the handover adds to the free evolution.
The idle of $152~\text{ns}$ multiplies the contrast by $0.901$ ($[0.871, 0.931]$).
We set this idle before the acquisition from a bench estimate of the handover, and it exceeds the $144~\text{ns}$ of the handover by $8~\text{ns}$.
The pre-specified analysis widens the intervals of the phase and of the contrast ratio for a mismatch of up to $50~\text{ns}$, to $[-0.013, 0.091]~\text{rad}$ and $[0.933, 1.078]$, which still lie within the margins of $\pm 0.10~\text{rad}$ and $0.90$ to $1.10$.
The drive phase is therefore continuous across the handover to within $0.1~\text{rad}$, and the RB survival of the split, unbroken and idling sequences differs by at most $0.006$ at $m = 16$ and $64$.

The cost of a handover inside a circuit is the decoherence during its idle of $144~\text{ns}$.
On this qubit, the idle multiplies the Ramsey contrast by $0.903$, which corresponds to an average gate infidelity of $(1 - 0.903)/3 \approx 0.03$ for pure dephasing~\cite{nielsen2002fidelity}.
The joint fit of the split and unbroken RB sequences gives a survival factor per handover of $0.987$ (95\% CI $[0.940, 1.049]$, or $[0.940, 1.048]$ when the floor $B$ is resampled as well).
The depolarizing model of the fit~\cite{magesan2011scalable} translates this factor into an error of at most $0.03$.
Carrying this error to long sequences requires the Markovian, time-independent noise model behind the RB fit, in which every handover multiplies the decay by the same factor independently of the others.
Under this model, with a handover about once per $1020$ Cliffords (one buffer half of $2048$ commands at about $2$ pulses per Clifford), the handovers add at most $0.03/1020 \approx 3 \times 10^{-5}$ to the fitted EPC, about $1\%$ of the EPC measured on this qubit.
The handovers also lengthen the sequence by $144~\text{ns}$ per about $71~\mu\text{s}$ of gates, or $0.20\%$.
This lengthening equals the relative bias of the EPC only if the error per Clifford comes from decoherence at the same rate during idle and driven evolution, which temporally correlated noise need not obey.

The $5101$-Clifford sequence compiles to $10207$ commands and runs its $2048$ shots in a median of $1804.08~\text{ms}$, against the programmed $1802.50~\text{ms}$.
Its five handovers per shot account for $5 \times 144~\text{ns} \times 2048 = 1.47~\text{ms}$ of the difference of $1.58~\text{ms}$, and the rest is the fixed cost of starting the run and draining its data.
The EPC comes from the fit $B + A p^m$ of the survival versus $m$~\cite{magesan2011scalable}, with the floor $B$ from the readout calibration, the amplitude $A$ and the decay $p$ per Clifford.
Its difference between the stacks over $m \le 512$, $+0.0003$, has a 95\% CI of $[-0.0001, 0.0007]$ from a bootstrap~\cite{efron1993bootstrap} over the $8$ random sequences, paired across the stacks.
This interval lies within the margin of $\pm 0.005$ and bounds the difference to less than a quarter of the EPC ($0.0031$ and $0.0034$).
These comparisons are between complete stacks, and the contribution of each \sys component is measured on the bench.

\section{Noise analysis}
\label{sec:app-noise}
A frequency offset $\delta f$ shifts the mean of the pair difference $d$ by $2K \cdot 2\pi \tau_R\, \delta f$.
Here $\tau_R = 0.5~\mu\text{s}$ is the free evolution, and $K$ is the slope of the Ramsey signal versus phase, calibrated at five phases after each record ($|K| = 0.22$ to $0.28$ per radian, standard error $0.03$).
Both stacks acquire the same $25000$ pairs, in $100.00~\text{s}$ on \sys and $100.31~\text{s}$ on \qubic.
A gap-aware estimator on timestamped runs could recover part of the lower frequencies from the baseline, so the comparison holds for this acquisition and analysis.
We compute the Hann-windowed periodogram~\cite{harris1978windows} of $d$, sampled once per pair (every $4~\text{ms}$, \ie, at $250~\text{Hz}$, with a Nyquist frequency of $125~\text{Hz}$).
We fit it from $0.1$ to $100~\text{Hz}$ with a white level plus a $1/f$ term seen through the pair response $\cos^2(\pi f \cdot 2~\text{ms})$, using the Whittle likelihood~\cite{whittle1953analysis}.
The likelihood ratio tests the $1/f$ term, with its null distribution from simulated records of binary outcomes.
The fits use the unbinned periodogram ordinates.
Every simulated record ($1000$ for the null distribution and $500$ per $1/f$ level for the upper limit) repeats the windowing and the fit, with $K$ drawn from its calibration uncertainty.

Both \sys records are consistent with white noise ($p = 0.39$ and $0.28$), with white levels $0.57\%$ below and $1.00\%$ above the binomial floor of the single-shot outcomes.
The simulations assume Gaussian $1/f$ noise from $0.01$ to $250~\text{Hz}$, generated at every measurement so that the folding of $125$ to $250~\text{Hz}$ into the analysis band is included.
They also assume a slope that stays at its calibrated value during the record.
Averaged over the calibration uncertainty, the $1/f$ level at which only $5\%$ of simulated records give a likelihood ratio at or below the measured one is $12.6$ and $17.7~\text{kHz}/\sqrt{\text{Hz}}$ at $1~\text{Hz}$ for the two records.
Their $95\%$ Monte Carlo intervals, from the binomial scatter of the simulated fractions, are $[12.1, 13.3]$ and $[17.1, 19.3]$.
The larger level, $\sqrt{S_f(1~\text{Hz})} \le 17.7~\text{kHz}/\sqrt{\text{Hz}}$, is a nominal $95\%$ upper limit.
Analysed in the $2.048~\text{s}$ runs of the baseline, the same records give $22.5$ and $25.3~\text{kHz}/\sqrt{\text{Hz}}$, and the two \qubic records give $22.3$ and $21.8~\text{kHz}/\sqrt{\text{Hz}}$.
Run by run, the \qubic records are also consistent with white noise above $0.49~\text{Hz}$ ($p \ge 0.07$).
The limits scale with the inverse of the calibrated slope, so a slope that drifted during a record by its standard error would move them by $12.30\%$ and $15.69\%$ for the two records.
Noise above $250~\text{Hz}$, which the simulations leave out, would fold into the analysis band as an almost white contribution, and the measured white levels bound this contribution.

In simulated records ($600$ without and $300$ per level with a $1/f$ component), the two whole \sys records detect a $1/f$ component with $50\%$ probability at a $5\%$ false-alarm rate from $\sqrt{S_f(1~\text{Hz})} = 10.5$ and $13.1~\text{kHz}/\sqrt{\text{Hz}}$.
The same records analysed in the $2.048~\text{s}$ runs of the baseline detect it from $12.3$ and $16.7~\text{kHz}/\sqrt{\text{Hz}}$.
The whole records thus detect spectral densities lower by factors of $(12.3/10.5)^2 = 1.37$ and $(16.7/13.1)^2 = 1.62$.
The $95\%$ Monte Carlo intervals of these factors, from the binomial scatter of the simulated detection rates, are $[1.22, 1.61]$ and $[1.39, 1.92]$, and the $5\%$ false-alarm threshold from $600$ null records carries $\pm 0.89\%$.
Below $0.49~\text{Hz}$, only the uninterrupted record measures the spectrum at all.
In a test that was not pre-registered, the correlation between the outcome differences of consecutive pairs, $4~\text{ms}$ apart, is $-0.001 \pm 0.003$ averaged over the four records, where the uncertainty is one standard error for white noise.
Quasi-static frequency noise that stays constant over several consecutive pairs and alone produced the dephasing ($T_2^* = 1.4~\mu\text{s}$) would correlate them by $0.127$ if the Ramsey signal decayed as a Gaussian and by $0.242$ if it decayed exponentially.

The acquisition departs from the pre-registered plan~\cite{antq} in four points.
First, it has two \sys records where the plan has at least three.
Second, each record has one slope calibration, taken after it and versus the phase of the second $\pi/2$ pulse, where the plan has one before and one after it, versus detuning.
Third, the free evolution is $0.5~\mu\text{s}$ instead of $0.3\,T_2^* = 0.42~\mu\text{s}$.
Fourth, no fast-cadence run checks for aliasing.
Because the host gaps of the baseline were not timestamped, its records are analysed per run only.
The analysis band ends at $100~\text{Hz}$, within the ``well below $125~\text{Hz}$'' of the plan, and the exponent of the $1/f$ term is fixed at $1$ because no component is resolved.

\section{Throughput measurement}
\label{sec:app-throughput}
The per-shot time of a circuit is the increase of its execution time on the \qubic baseline bitfile from $100$ to $50100$ shots, divided by $50000$.
Both runs use the \sys Uplink of this bitfile.
All $50100$ shots, with their two readout records per qubit, are thus kept in one uninterrupted run whose readout transfer overlaps with execution, and the increase is the execution time of the added shots.
For \sys, which runs on its own bitfile, $\Delta$ also contains the per-shot difference between the two bitfiles.
Over the five runs of a circuit on \sys, the standard deviation of the execution time is at most $0.09~\text{ms}$ (median $3.5~\mu\text{s}$).
Combined with the uncertainty of the per-shot time, the standard error of $\Delta$ is at most $0.04~\text{ms}$ for the circuits with at least $1000$ shots.

The \sys Downlink combines three mechanisms, which the measurements do not separate.
It overlaps the circuit transfer with execution and readout transfer, omits the commands of unused cores and transfers circuits by DMA through DRAM.
To separate the fixed cost of a circuit from the per-shot difference, we time every stage on the PS in a separate series of runs (Table~\ref{tab:decomp}).
On the \sys bitfile, a shot takes its compiled length plus $150$ to $156~\text{ns}$, and on the baseline bitfile its compiled length plus $141$ to $170~\text{ns}$.
The reference time can therefore be longer or shorter than the execution on \sys, depending on the circuit.
For the Bell state, a shot is $10~\text{ns}$ longer on the baseline bitfile, which adds up to $0.25~\text{ms}$ over $24576$ shots and makes the $\Delta$ of \sys $-0.21~\text{ms}$.
The time outside the execution of \sys, \ie, uploading the command images, starting, launching and draining the last readout data, is $38.3$ to $220.9~\mu\text{s}$ per circuit (median $74.3~\mu\text{s}$).
This time is at most $0.48\%$ of the execution time of \sys for the circuits with at least $1000$ shots.
The drain stays below $30~\mu\text{s}$ in all but $9$ of the $220$ runs and takes at most $0.75~\text{ms}$.
This fixed cost becomes noticeable only for GHZ-6 and GHZ-8, whose reference times are $2.72$ and $0.88~\text{ms}$.

\input{tables/decomp}

The batches run in five blocks, each holding two of the ten random batches, and every batch is measured five times in its block.
The 95\% CIs of the throughput gain are $t$ intervals over the five block means, so they reflect the variation between batch compositions.
The ten batches give gains of $52.71\%$ to $72.10\%$ against the stock baseline and $44.84\%$ to $64.59\%$ against the baseline with the shared table layout.
The five measurements of a batch have relative standard deviations of at most $0.09\%$ on the baseline and below $0.01\%$ on \sys.
Our measurement plan designated the comparison against the baseline with the shared table layout (Section~\ref{sec:design-overview}), an option of the \qubic assembler, as confirmatory.
The two gains differ by this layout.
With it, the baseline runs the random batches in $3794.18~\text{ms}$ on average instead of $4009.81~\text{ms}$ (Table~\ref{tab:batch_speedup}).

\section{Early-termination statistics}
\label{sec:app-stop}
Every circuit has enough shots for the grading.
For all $20$ circuits and at every checkpoint from the third one on, $200$ independent noiseless samples of the shot count of the checkpoint give a $95$th percentile of the TVD below $0.3$ and a $5$th percentile of the HF above $0.7$.
Over the $2160$ runs of the $20$ seeds, the rule stops one of the $117$ \textsc{Fail} runs only at the last checkpoint.
The $3$ stopped \textsc{Marginal} runs are a GHZ-3 run whose TVD and HF stay within $0.006$ of the \textsc{Fail} boundary throughout and two GHZ-8 runs, which stop when only $12$ and $15$ of their $32$ shots have been taken.
In $100$ further seeds of every circuit on every backend that can hold it (seeds $20$ to $119$, $10800$ runs), the rule stops all $588$ \textsc{Fail} runs (one of them only at the last checkpoint), $12$ of the $1438$ \textsc{Marginal} runs and none of the $8774$ \textsc{Pass} runs.
Over all $120$ seeds, it thus stops a \textsc{Pass} run of these circuits on these simulated backends with a probability below $0.04\%$ ($95\%$ Clopper--Pearson bound~\cite{clopper1934confidence}, $0$ of $10537$).
Because the rule decides on prefixes of the same shots from which the full-budget grade is computed, this rate measures agreement with the full-budget grade of each simulated run.
To test the rule against a reference that shares no shots with the run it decides on, we use the full-budget grade of another seed of the same circuit and backend.
Paired this way (seed $s$ with seed $s + 60$ modulo $120$), the rule stops none of the $10537$ runs whose reference is graded \textsc{Pass}.
Over all pairs of seeds of a circuit and backend, a stopped run meets a \textsc{Pass} reference in $30$ of the $1253903$ pairs with a \textsc{Pass} reference.
All $30$ come from GHZ-8 on FakeTorino, where $5$ of the $120$ seeds are graded \textsc{Pass} and $115$ \textsc{Marginal} and the rule stops $6$ runs.
The decisions depend little on the thresholds of the rule.
Over the $27$ combinations of $\mathrm{HF} \le 0.35$, $0.40$ or $0.45$, $\mathrm{TVD} \ge 0.55$, $0.60$ or $0.65$, and $\epsilon_{\mathrm{TVD}} = \epsilon_{\mathrm{HF}} = 0.02$, $0.03$ or $0.05$, no \textsc{Pass} run is stopped.
Across these combinations, the rule stops between $585$ and all $705$ \textsc{Fail} runs and between $6$ and $245$ of the $1718$ \textsc{Marginal} runs.

On the board, the shots executed after the stop point are a larger fraction of the shot budget for circuits with fewer shots, because the same feedback delay costs a larger fraction of a smaller budget.
The delay depends on the checkpoint computation of each circuit and on how often the PS polls for new readout data.
The preset stop points can also differ from the simulated ones.
For GHZ-8, the preset checkpoint $4$ lies below the median stop checkpoint $6$ of the simulation.
We therefore combine the stop checkpoints of all $120$ seeds with the per-shot time and the mean stop latency measured on the board.
Over the stopped runs, a stop then saves a median of $69.69\%$ (GHZ-3), $50.00\%$ (GHZ-6), $63.90\%$ (QML Image Class), $69.38\%$ (VQE BeH$_2$), $69.68\%$ (VQE SrH PDM) and $21.77\%$ (GHZ-8) of the execution time of the full circuit.
With the largest latency measured, a stop comes too late to save any time in $6$ of the $126$ stopped GHZ-8 runs and in the one GHZ-3 run that stops at its last checkpoint.

%% file: tables/handover.tex
\begin{table*}[t]
\centering
\caption{Events of one command-buffer handover in the DSP clock domain ($2~\text{ns}$ per cycle), from the cycle-accurate simulation~\cite{antq}, counted from the last sample of circuit $k$ at the DAC interface. The next halves must be ready, with their fill state synchronized into the DSP clock domain, before the halves of circuit $k$ are released, and every later step is fixed. The gap counts the idle cycles strictly between the two samples, one less than the difference of their cycle numbers.}
\label{tab:handover}
\setlength{\aboverulesep}{0pt}
\setlength{\belowrulesep}{0pt}
\renewcommand{\arraystretch}{1.3}
\footnotesize
\begin{tabularx}{\textwidth}{>{\raggedright\arraybackslash}X|c}
\toprule
\rule{0pt}{2.5ex}Event & Cycle \\
\midrule
Last sample of circuit $k$ leaves the signal pipeline & $0$ \\
Signal pipeline reports idle & $1$ \\
Program of circuit $k$ ends & $3$ \\
DSP reports the last shot & $4$ \\
Sequencer releases the halves of circuit $k$, and the next halves must be ready & $6$ \\
Next halves seen ready (the wait counted by the flag is $1$ cycle when they were ready before) & $8$ \\
Sequencer starts circuit $k+1$ on all cores & $10$ \\
First sample of circuit $k+1$, at its programmed start time plus $58$ cycles of processor start and element and DAC pipeline & $10 + T_0 + 58$ \\
\midrule
Handover, \ie, gap minus programmed start time $T_0$ ($60$ cycles between two shots of one program) & $67$ ($134~\text{ns}$) \\
\bottomrule
\end{tabularx}
\end{table*}

%% file: tables/decomp.tex
\begin{table*}[t]
\centering
\caption{Time of a single circuit outside its execution on the \sys bitfile, from the stage timestamps of the PS (mean of $11$ runs per circuit, one warm-up included). The execution itself takes the compiled shot length plus $150$ to $156~\text{ns}$ per shot.}
\label{tab:decomp}
\setlength{\aboverulesep}{0pt}
\setlength{\belowrulesep}{0pt}
\renewcommand{\arraystretch}{1.3}
\footnotesize
\begin{tabularx}{\textwidth}{l|>{\raggedright\arraybackslash}X|c|c}
\toprule
\rule{0pt}{2.5ex}Stage & Contents & \makecell{Median\\$(\mu\text{s})$} & \makecell{Range over the $20$\\circuits $(\mu\text{s})$} \\
\midrule
Upload & DMA of the command images from the PS to DRAM & $30.5$ & $8.2$--$90.6$ ($1$ to $14$ cores) \\
Start & configuration and start written by the PS & $0.7$ & $0.7$--$0.8$ \\
Launch & first fetch into the command buffer and start of the DSP, as seen by the PS & $18.5$ & $6.2$--$53.4$ \\
Drain & last readout data from DRAM into PS memory & $16.1$ & $13.0$--$94.7$ \\
\midrule
Total & the time outside the execution & $74.3$ & $38.3$--$220.9$ \\
\bottomrule
\end{tabularx}
\end{table*}

%% file: sections/Acknowledgement.tex
\ack{This work is supported by a collaboration between the US DOE and the National Science Foundation (NSF).
This material is based upon work supported by the U.S. Department of Energy, Office of Science, National Quantum Information Science Research Centers, Quantum Systems Accelerator (Award No. DE-SCL0000121).
Additional support is acknowledged from NSF Safe-OSE program (Award No. 2533222).}